\documentclass{article}
\usepackage{PRIMEarxiv}
\usepackage[utf8]{inputenc} 
\usepackage[T1]{fontenc}    
\usepackage{textcomp}       
\usepackage{hyperref}       
\usepackage{url}            
\usepackage{booktabs}       
\usepackage{amsfonts}       
\usepackage{amsmath}        
\usepackage{amssymb}        
\usepackage{nicefrac}       
\usepackage{microtype}      
\usepackage{fancyhdr}       
\usepackage{graphicx}       
\usepackage{tabularx}       
\usepackage{subcaption}     
\usepackage{needspace}
\usepackage{tabularx}
\usepackage{booktabs} 
\usepackage{authblk}
\usepackage{float}
\usepackage{placeins}
\usepackage{array}
\usepackage[style=authoryear, natbib=true, backend=biber]{biblatex}
\usepackage{enumitem}
\setlist{nosep} 

\graphicspath{{media/}}     

\title{Trustworthy AI Posture (TAIP): A Framework for Continuous AI Assurance of Agentic Systems at Horizontal and Vertical scale 
\thanks{\textit{\underline{Citation}}: 
\textbf{Lupo, Vo, Locke. Trustworthy AI Posture (TAIP): A Framework for Continuous Assurance of Agentic AI Systems. arXiv preprint (January 2026).}} 
}

\author[1]{Guy Lupo, PhD Researcher}
\author[1]{Bao Quoc Vo, Associate Professor}

\author[2]{Natania Locke, Associate Professor and Dean}

\affil[1]{School of Science, Computing and Emerging Technologies}

\affil[2]{Swinburne Law School}

\affil[ ]{Swinburne University of Technology, Melbourne, Australia}

\affil[ ]{\textit{\{glupo, bvo, nlocke\}@swin.edu.au}}

\date{} 

\DeclareUnicodeCharacter{2121}{}
\usepackage{newunicodechar}
\newunicodechar{τ}{\ensuremath{\tau}}
\newunicodechar{≥}{\ensuremath{\geq}}

\begin{document}
\maketitle
\begin{center}
\textit{February 2026}
\end{center}

\begin{abstract}
The emergence of self-coding, high-velocity, ephemeral Agentic AI in organisations is creating an internal audit and assurance scalability crisis. Point-in-time, document-based audits cannot keep pace with non-deterministic behaviour and distributed deployments in rapidly evolving Agentic AI environments. The crisis is twofold: vertically, existing frameworks struggle to adapt to rapidly evolving governance and control environments; horizontally, they cannot scale across increasingly complex, distributed systems and heterogeneous evidence sources. Risk-based regulation now requires organisations to demonstrate trustworthiness through ongoing control adequacy and effectiveness, yet current Trustworthy AI assurance frameworks remain fragmented and largely manual.
A useful precedent exists in cybersecurity. As threats and systems scaled, security assurance evolved from periodic checklist compliance toward continuous posture management, where control telemetry is monitored against risk thresholds to sustain trust at scale. This paper applies the same structural insight to Trustworthy AI assurance: trustworthiness must be maintained as a continuously generated signal rather than a static certificate.
This paper contributes (1) a Trustworthy AI Assurance Ontology that integrates governance, operations, and audit handoffs including prior partial ontologies into an end-to-end trust-signal pathway from regulatory obligation to verifiable evidence; (2) an ontology-driven comparative analysis and evidence-gated benchmark of thirteen leading frameworks, revealing a structural posture-readiness gap; and (3) the Trustworthy AI Posture (TAIP) framework, which operationalises the NIST AI RMF Test–Evaluate–Verify–Validate cycle as reusable AI Assurance Objects.
TAIP decouples policy checklists (what) from execution semantics (how), enabling the same assurance mechanics to be instantiated and automated across jurisdictions and alongside ephemeral agents. Evidence from heterogeneous tools is normalised into success/fail outcomes and recursively aggregated into posture at claim, system, organisational, and ecosystem levels. A work-in-use case mapping Australian AI Guardrails to Microsoft 365 Copilot demonstrates claim decomposition, evidence binding, and posture determination in practice. By standardising execution while allowing policy variation, TAIP addresses both the vertical governance challenge and the horizontal technological scalability challenge, enabling trust signals to operate at machine speed.
\end{abstract}



\section{Introduction: The Agentic AI Assurance Scalability Crisis}
\label{sec:intro}
Agentic AI systems are rapidly becoming foundational infrastructure for modern organisations, autonomously orchestrating tools, generating decisions, and executing complex workflows across multiple environment at unprecedented scale. As AI systems become deeply integrated into the societal infrastructure, ensuring public trust is no longer optional, but a prerequisite for sustainable adoption. The industry currently faces a critical "Assurance Scalability Crisis": the volume, velocity, risks and complexity of modern AI deployments have outpaced the capacity of traditional, human-centric audit practices to govern them \citep{minkkinen_continuous_2022,ojewale_towards_2024,raji_closing_2020,jaffri_hype_2023}.

This crisis reaches a breaking point with the transition from static models to autonomous, tool-using, multi-agent systems \citep{schick_toolformer_2023,wu_autogen_2023,liu_agentbench_2023,jimenez_swe-bench_2024,li_camel_2023,wang_voyager_2023}. As evaluation shifts toward interactive agent benchmarks, human-speed, document-based audits are physically incapable of monitoring machine-speed, ephemeral agents operating across complex tool chains. Traditional audits, reliant on retrospective checklists and episodic reviews, cannot capture the non-linear reasoning and tool-interaction loops inherent in Agentic workflows \citep{li_trustworthy_2023}. Without a fundamental shift in assurance methodology, organisations face not only operational risks but also systemic threats to regulatory legitimacy and enterprise accountability.

A similar scalability crisis was previously observed in cybersecurity, where static compliance audits failed to address high-velocity threat environments. The industry responded by shifting toward continuous posture management, allowing real-time visibility into control effectiveness and behavioural anomalies rather than periodic certification. AI assurance has not yet undergone an equivalent structural transition.

\subsection{The Shift to Risk-Based Regulation and the governance gap}

Regulators across the EU, APAC, and parts of the United States are shifting toward risk-based regulation, recognising that rigid, prescriptive rules cannot keep pace with AI innovation and may stifle development. The EU Artificial Intelligence Act (AIA) \citep{noauthor_regulation_2024}, for example, places responsibility on organisations to define, manage, and evidence risks according to their specific operational context. This transition signals a move away from checklist compliance toward context-informed, risk-informed accountability.
However, despite this regulatory evolution, a structural gap remains. Organisations are expected to adapt their control environments to rapidly evolving AI systems while simultaneously supporting accelerating innovation cycles. Existing evaluation practices remain largely conceptual or reliant on manual questionnaires such as ALTAI \citep{comission_assessment_2020}. Operational evidence including logs and telemetry   is rarely systematically linked to governance claims \citep{brundage_toward_2020,haugen_assurance_2025,batarseh_ai_2023}. As a result, demonstrating adequate, effective and sustainable controls \citep{lawson_survey_2017,calagna_applying_2021,roberta_provasi_updated_2015} remains inconsistent and resource-intensive.
At the same time, the governance landscape continues to expand vertically, with new AI-specific risks and regulatory obligations. Horizontally, AI deployments are becoming more distributed, tool-driven, and agentic. Traditional point-in-time audits, conducted as retrospective and project-based exercises \citep{li_making_2024}, cannot keep pace with this dual expansion. In agentic environments, each system or agent effectively requires separate review cycles \citep{minkkinen_continuous_2022,luciano_capai-procedure_2022,javadi_monitoring_2021,radanliev_operationalising_2026}.
Trustworthy AI assurance therefore fails to scale. There are insufficient auditors, insufficient time, and insufficient accessible structured evidence to provide assurance at the speed and volume required by machine-driven systems.

\subsection{Problem Framing and Research Scope}
The persistence of this gap, despite the risk-based regulatory reform, indicates that the constraint is architectural rather than regulatory. AI assurance remains organised around document-centric workflows and episodic review cycles, while agentic systems operate at machine speed, across distributed environments, and within structurally complex toolchains.
For Trustworthy AI Assurance to function on a comparable scale, similar to the evolution of cybersecurity posture management, it must address two distinct but interrelated structural challenges.
\paragraph{The Vertical scale - fast evolving governance and control environment}

Governance, risk, and compliance (GRC) environments are not static. New AI-specific risks, regulatory obligations, and control expectations continually shape the control landscape. Assurance must therefore support continuous control monitoring , not just verifying that controls exist but demonstrating that they are adequate (fit for purpose within a defined risk context), effective (supported by operational evidence that they function as intended) and sustainable (maintained over time without prohibitive cost or excessive manual intervention)\citep{calagna_applying_2021,isaca_cobit_2018}. Vertical traceability requires the structured ability to bind evolving governance intent to operational controls and their associated evidence across diverse control configurations.
\paragraph{The Horizontal scale - Multiple technology environment}

AI systems are evolving at unprecedented speed. Agentic systems can autonomously execute tasks, orchestrate tools, spawn instances, and exhibit non-deterministic behaviour. The rate of technological innovation expands continuously across heterogeneous technical environments. Assurance mechanisms must therefore scale horizontally not only across multiple deployments, but across dynamically changing architectures capable of autonomous execution and replication.

Episodic project-based audits cannot keep up with machine-speed AI systems. A different structural model is required. Section \ref{sec:state-of-the-art-lit-review} addresses this challenge through a structured literature review and comparative analysis. An ontological analysis is used to systematically identify and examine existing Trustworthy AI audit and assurance frameworks. The objective is to determine whether the current state of the art provides the structural foundations necessary for a posture-based transition comparable to that achieved in cybersecurity.
The analysis reveals clear structural gaps in existing approaches. These findings lead directly to the formulation of the research question \ref{content:RQ1}. The question focuses on what Trustworthy AI requires to operate as a scalable posture system. This includes identifying the essential components, processes, and architectural conditions needed to bind governance intent to operational evidence at machine speed.
Section \ref{sec:engineering-taip} then presents the development and validation of the Trustworthy AI Posture (TAIP) framework as a structural response to these requirements.

\subsection{Research Methodology, Innovation, and Contributions}
This paper primary contribution is the engineering of Trustworthy AI assurance as an executable, recursively composable signal-generation architecture capable of operating at both governance (vertical) and technological (horizontal) scale \citep{tabassi_artificial_2023,minkkinen_continuous_2022, raji_anatomy_2024}. Existing assurance practices remain largely checklist-driven and retrospective \citep{donati_beyond_2025}. In contrast, this work reframes trustworthiness as a continuously generated and validated signal, analogous to cybersecurity posture management \citep{cheruku_ai-driven_2025,al-karaki_gosafe_2022}. By shifting the unit of analysis from the “report” to the “signal,” trustworthiness becomes a dynamic output that must be observed, tested, and sustained in real time.
The methodological progression followed three stages: empirical diagnosis, structural formalisation, and engineering response. Each stage builds directly on the previous one.
\paragraph{Contribution \#1 Section(\ref{sec:taio-signal})}:
 introduces the Trustworthy AI Assurance Ontology as a formal model of how trustworthiness is generated. Developed through inductive mapping of existing assurance practices, the ontology reconstructs the end-to-end pathway linking governance intent to operational evidence. Unlike taxonomy-style Trustworthy AI ontologies, which describe what trustworthiness is, this model captures how trustworthiness is operationalised. It integrates regulation, accountable management, governance, risk, controls, operations, and audit into a single closed-loop signal architecture.
\paragraph{Contribution \#2 (Section \ref{subsec:rubric-gate})}:
 operationalises this structure to conduct an ontology-driven comparative analysis of thirteen leading AI audit and assurance frameworks. Using structured data extraction and a rubric-gated benchmark, the study evaluates posture-readiness   defined as the structural ability to support continuous signal-based assurance at both vertical (GRC traceability) and horizontal (technological scalability) levels. The analysis reveals a systemic gap: existing frameworks remain document-centric and lack automation and continuous monitoring capabilities required for agentic, non-deterministic environments.
\paragraph{Contribution \#3 Section (\ref{subsec:taip-at-a-glance})}:
 presents the Trustworthy AI Posture (TAIP) framework as a structural response to this gap. TAIP encapsulates assurance logic as an AI Assurance Object implemented using a Composite Design Pattern \citep{gamma_elements_1995}. The object embeds TEVV (Test–Evaluate–Verify–Validate) as an executable assurance unit and supports recursive aggregation through a Posture Tree, enabling vertical alignment with Governance, Risk, and Compliance structures \citep{isoiec_iso_2020,isaca_cobit_2018} and binding at the Operational Design Domain boundary \citep{herrera-poyatos_responsible_2025}.
TAIP was evaluated using the same ontology-driven benchmark and achieved Level 4 (“posture-ready”). Its practical viability was demonstrated by mapping the Australian Government’s AI Guardrails to a Microsoft 365 Copilot deployment. Together, the benchmark evaluation and applied validation confirm TAIP’s architectural soundness and operational readiness for scalable, agentic AI environments.

\section{State of the Art, Gap Analysis, Identifying the Trustworthiness Signal, and the need for TAI Posture}
\label{sec:state-of-the-art-lit-review}

The transition from episodic auditing to continuous trustworthiness signals requires rethinking how governance intent, operational controls, and audit validation are structurally connected. While the literature provides rich taxonomies of AI ethical principles and domain specific technical metrics \citep{attard-frost_ethics_2023,jobin_global_2019,kaur_trustworthy_2023}, it lacks a unified structural flow mechanism to bind high-level governance claims to low-level operational evidence at scale.
In cybersecurity, continuous monitoring and assurance of the adequacy, effectiveness, and sustainability of controls \citep{calagna_applying_2021} throughout the lifecycle is also known as a "posture" \citep{al-karaki_gosafe_2022}. The concept of posture provides a holistic view of an organisation's security and compliance status by continuously assessing the effectiveness of controls and validating them against company policies and regulatory obligations.
Applying posture to AI systems enables trustworthiness to be continuously assessed throughout development, deployment, operation, and retirement. Instead of producing a retrospective report, posture provides a measurable and ongoing performance indicator \citep{javadi_monitoring_2020}.
To examine whether such a posture model is feasible for Trustworthy AI, Section \ref{sec:taio-signal} conducts an inductive analysis of existing frameworks and practices. This analysis identifies the core components involved in assurance and maps the flow through which trustworthiness is generated and resulted in the creation of the Trustworthy AI Assurance Ontology.
Section \ref{subsec:ontology-driven-lit-analysis} then applies a comparative analysis using a rubric-gated benchmark (\ref{subsec:rubric-gate}) to evaluate current approaches. The findings expose structural gaps, which are synthesised into the research questions presented in \ref{content:RQ1}.

\subsection{The Development of the Trustworthy AI Assurance Ontology: Identifying the Trustworthiness Signal}
\label{sec:taio-signal}

In the context of AI Governance Risk and Compliance, trustworthiness is defined as \emph{a validated and verified claim of responsible use of AI technology} \citep{isoiec_iso_2022}. A verified claim forms a single atomic unit of trustworthiness. Trustworthiness therefore depends on repeatable and auditable evidence, rather than periodic self-attestation or adherence to a checklist \citep{donati_beyond_2025}.

\paragraph{From ``trustworthy AI'' taxonomies to an assurance-flow ontology.}
Most ``Trustworthy AI'' ontologies organise principles, requirements, and conceptual relationships, or model trust as a stakeholder concept \citep{harrison_ontology_2021,amaral_ontology-based_2020}. They support shared language, but they rarely model the operational handovers that turn governance intent into tested evidence. In practice, this gap encourages static checklist-style compliance.

The Trustworthy AI Assurance Ontology has a different purpose. It does not define what trustworthiness \emph{is}. It defines how trustworthiness is \emph{demonstrated}. It captures the process logic and dependencies that produce a repeatable \emph{trustworthiness signal}. It is therefore an ontology of assurance \emph{flow}, not a taxonomy of trust concepts.

\subsubsection{Empirical grounding and foundations in prior ontology work}
The Trustworthy AI Assurance ontology was grounded in observed assurance practices. The review included operational evidence and best-practice documents across regulation, accountable management, AI GRC, AI operations, audit, and frameworks. Based on the review the recurring relationships were modelled between these entities by tracing the exchanged (claims, policies, risks, controls, evidence, findings) and how feedback returns to governance.

Existing ontologies were also included in the review process to test whether an end-to-end model already existed. Strong yet \emph{partial} ontologies were found that cover key slices of the process:
\begin{itemize}
    \item \textbf{Trust concepts and trust signals:} \citet{harrison_ontology_2021} structures trustworthiness concepts, while \citet{amaral_ontology-based_2020} relates trustworthiness requirements to trust-warranting signals.
    \item \textbf{Accountability-to-governance traceability:} SAO and RAInS connect accountability plans to governance so evidence capture is planned and readable \citep{verborgh_semantic_2021, naja_using_2022}.
    \item \textbf{Testing workflow structure:} AI-T formalises testing activities and their sequence for AI-based systems \citep{olszewska_ai-t_2020}.
    \item \textbf{Risk representation:} AIRO encodes AI risks in a machine-readable form aligned to regulatory and ISO risk management concepts \citep{golpayegani_airo_2022}.
    \item \textbf{Impact assessment artefacts:} FRIA structures fundamental rights impact assessments so compliance artefacts become computable \citep{rintamaki_developing_2024}.
    \item \textbf{Operational provenance over time:} AIBOM extends SBOM ideas to track changing AI components (e.g., retraining, drift, agentic decision flows) to support lifecycle evidence integrity \citep{radanliev_operationalising_2026}.
\end{itemize}

These works show that assurance elements can be modelled as operational entities and dependencies. The remaining gap is end-to-end integration. The reviewed ontologies are largely domain-scoped, and they do not connect governance intent, operational controls, evidence generation, and independent audit evaluation into one continuous pathway.

\begin{figure}[H]
    \centering
    \includegraphics[width=0.75\linewidth]{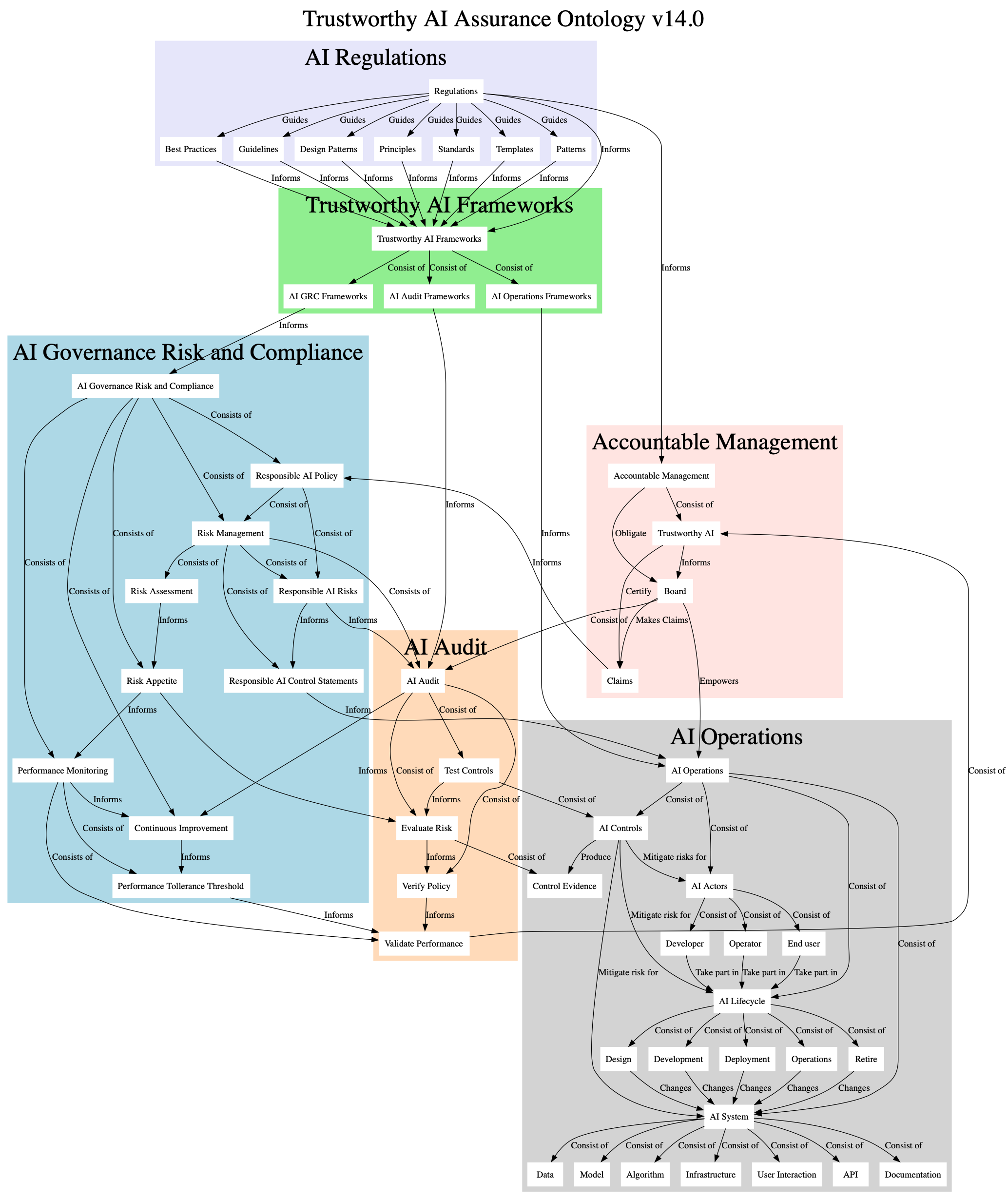}
    \caption{Trustworthy AI Assurance Ontology (diagram).}
    \label{fig:Figure1}
\end{figure}
\FloatBarrier

\subsubsection{Formalising the trustworthiness signal and analysing scale}
Following the review outcomes the observed workflow was formed into the Trustworthy AI Assurance Ontology (Figure~\ref{fig:Figure1}). The ontology consolidates six assurance domains into one semantic architecture and makes the closed-loop assurance pathway explicit.

\begin{figure}[H]
    \centering
    \includegraphics[width=1\linewidth]{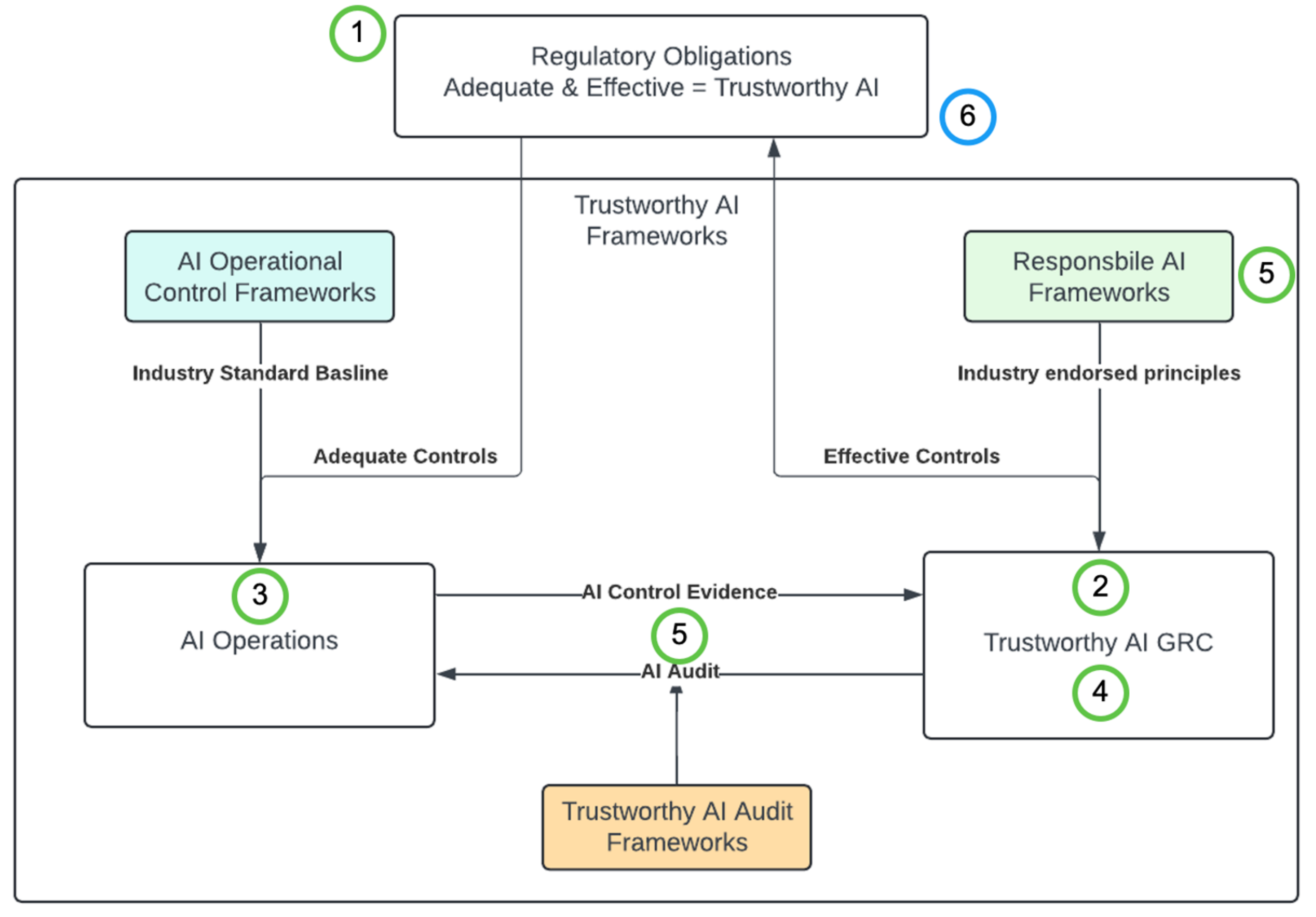}
    \caption{Establishing trust in AI systems: assurance domains and signal flows.}
    \label{fig:Figure2}
\end{figure}

Operationally, the trustworthiness signal is the repeatable pathway (Figure~\ref{fig:Figure2}):
\[
\text{Regulation} \rightarrow \text{Claim} \rightarrow \text{Risk/Policy} \rightarrow \text{Control} \rightarrow \text{Evidence} \rightarrow \text{Audit} \rightarrow \text{Governance feedback}.
\]
This supports vertical scaling because low-level control failures can invalidate the specific governance claim they support \citep{mokander_us_2022}. It also supports continuous assurance because evidence and testing cadence can be aligned with risk and structured as TEVV activity \citep{national_institute_of_standards_and_technology_artificial_2023}. Full node definitions and relationships are provided in Appendix~A.

This result is consistent with current traditional audit and assurance practices \citep{minkkinen_continuous_2022, mokander_auditing_2024}. Trustworthy AI audit frameworks shape how organisations work by defining methods, artefacts, and assurance routines \citep{li_trustworthy_2023, lu_responsible_2024,baker-brunnbauer_taii_2023}. When these frameworks are built on checklist-based, deterministic logic and rely on point-in-time reviews, they reproduce the periodic and non-scalable assurance model observed today. Identifying audit frameworks as the primary leverage point therefore provides a clear direction for the following literature review and comparative analysis, focusing attention on how these frameworks must evolve to enable scalable and automatable AI trustworthiness.

\subsection{Ontology-Driven Literature Review and Comparative Analysis}
\label {subsec:ontology-driven-lit-analysis}
With Trustworthy AI audit frameworks identified as the principal leverage point for scalability, the next step was to examine whether the current state of the art is structurally capable of supporting continuous, signal-based assurance. The Trustworthy AI Assurance Ontology was operationalised as an analytical instrument to guide the identification and evaluation of relevant frameworks. By design, the review was constructed to generate an empirical evidence base through structured data extraction. This evidence was then used to develop a rubric-gated Posture-Readiness Benchmark, which underpins the comparative analysis. The process was conducted in accordance with the PRISMA guidelines to ensure transparency and reproducibility, but its primary purpose was analytical: to build a defensible, evidence-based classification framework grounded in the extracted data.
\begin{figure}[H]
    \centering
    \includegraphics[width=0.5\linewidth]{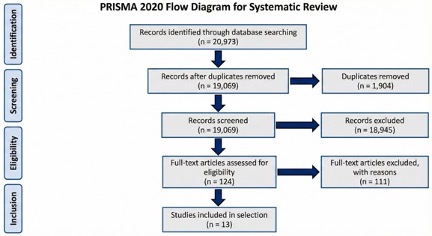}
    \caption{PRISMA 2020 flow diagram for the ontology-informed literature review (identification, screening, eligibility, inclusion).}
    \label{fig:PRISMA-Figures}
\end{figure}
\FloatBarrier

\begin{table}[H]
  \centering
  \small 
  \caption{Overview of the thirteen Trustworthy AI Audit and Assurance frameworks identified in the review.}
  \label{13-framework-table}
  
  \begin{tabular}{@{} p{3cm} p{1.2cm} p{3.5cm} p{6cm} @{}}
    \toprule
    \textbf{Framework} & \textbf{Year} & \textbf{Reference} & \textbf{Authors} \\
    \midrule
    Ethics-Based Audits & 2021 & \citep{mokander_ethics-based_2021-1} & Jakob Mökander and Luciano Floridi \\
    \addlinespace
    TAII & 2023 & \citep{baker-brunnbauer_taii_2023} & Josef Baker-Brunnbauer \\
    \addlinespace
    SMACTR & 2020 & Raji, I. D. et al. (2020) & Inioluwa Deborah Raji, Andrew Smart, Rebecca N. White, Margaret Mitchell, Timnit Gebru, Ben Hutchinson, Jamila Smith-Loud, Daniel Theron, and Parker Barnes \\
    \addlinespace
    GAFAI & 2022 & \citep{markert_gafai_2022} & Thora Markert, Fabian Langer, and Vasilios Danos \\
    \addlinespace
    CapAI & 2022 & Floridi, L. and Cowls, J. (2022) & Luciano Floridi and Josh Cowls \\
    \addlinespace
    Z-Inspection\textregistered & 2021 & \citep{zicari_z-inspection_2021} & Roberto V. Zicari, et al. \\
    \addlinespace
    ALTAI & 2020 & \citep{commission_assessment_2020} & High-Level Expert Group on AI (AI HLEG) \\
    \addlinespace
    AI Risk Management Framework & 2021 & \citep{national_institute_of_standards_and_technology_artificial_2023} & Elham Tabassi (NIST) \\
    \addlinespace
    ITAF & 2020 & \citep{noauthor_it_2020} & ISACA \\
    \addlinespace
    Audit and Assurance of AI Algorithms & 2021 & \citep{akula_audit_2021} & Ramya Akula and Ivan I. Garibay \\
    \addlinespace
    SAO Framework & 2021 & \citep{verborgh_semantic_2021} & Iman Naja, Milan Markovic, Peter Edwards, and Caitlin Cottrill \\
    \addlinespace
    Assurance Audits of Algorithmic Systems & 2024 & \citep{lam_framework_2024} & Khoa Lam, Benjamin Lange, et al. \\
    \addlinespace
    Hybrid-DLT Framework & 2024 & \citep{pelosi_hybrid-dlt_2023} & Andrea Pelosi, Claudio Felicioli, et al. \\
    \bottomrule
  \end{tabular}
\end{table}
\FloatBarrier

\subsubsection{Literature Review - Trustworthy AI Assurance and Audit Frameworks }

The review was guided by the following question, derived from the PICO-based framing used for the AI Audit Framework cluster:

\textbf{To what extent do existing Trustworthy AI Audit and ssurance frameworks remain fit for purpose in the large-scale, Agentic AI era?}

Where “fit for purpose” refers to structural scalability in the context of large-scale \citep{sevilla_compute_2022, mokander_auditing_2024, jaffri_hype_2023}, agentic AI systems. It describes a framework’s ability to define evidence artefacts, execution logic, and monitoring cadence in ways that can be automated and composed reliably at scale.

The search (Figure \ref{fig:PRISMA-Figures}) identified 20,973 (not included snowballing) records using a variety of keyword combination based on the Trustworthy AI Assurance Ontology - AI Audit Framework cluster, and its adjacent nodes \ref{sec:apdx:a:TAI-ontology}, with focus on the following venues ACM, IEEE, Springer, and Scopus.

After removing 1,904 duplicates, 19,069 records were screened. 18,945 were excluded at the screening, leaving 124 full-text articles assessed for eligibility; 111 were excluded mainly because they referred to the use of AI in Audit instead of defining Audit for AI, resulting in the following 13 relevant Trustworthy AI Audit Frameworks included for further analysis and benchmarking in Table \ref{13-framework-table}.

The review inclusion criteria deliberately focuses on frameworks that claim to support audit and assurance of AI systems (not only high-level ethics principles) with different ontological coverage percentage to ensure all aspect of production of trustworthiness signal are addressed.

This inclusion set represents the diverse landscape of current assurance practices, ranging from high-level policy instruments like the NIST AI Risk Management Framework (AI RMF 1.0)\citep{tabassi_ai_2023} and the EU's Assessment List for Trustworthy AI (ALTAI)\citep{commission_assessment_2020}, to academic methodologies such as Ethics-Based Auditing \citep{mokander_ethics-based_2021-1} and technical proposals like the Hybrid-DLT Based Trustworthy AI Framework \citep{pelosi_hybrid-dlt_2023}. The full detailed extractions for each framework can be found in Appendix B \ref{sec:apdx:b:frameworks}.

\subsubsection{Constructing an Evidence-First Scalability Capability Ladder for Comparative Analysis}
\label{subsec:rubric-gate}

To move beyond subjective maturity models based on self-assessment, an evidence-gated capability ladder was developed (see Appendix \ref{sec:apdx:b:frameworks} for the full rubric) to evaluate the structural readiness of existing frameworks. The ladder operationalises the Trustworthy AI Assurance Ontology as an assessment instrument across three interdependent pillars: Governance (G), Operations (O), and Audit (A).
Its construction followed two steps. First, the ontology was used to identify the structural requirements for operating at large scale across these clusters , reflecting the realities of Agentic AI systems \citep{jaffri_hype_2023, szarmach_model_2026}, including high-velocity AI operations \citep{li_trustworthy_2023, lu_responsible_2024, radanliev_operationalising_2026, solanki_operationalising_2023, haugen_assurance_2025},  , non-deterministic behaviour \citep{ mokander_auditing_2024}, and complex toolchains\citep{sevilla_compute_2022,, kpmg_ai_2025, agarwal_seven-layer_2024}. Second, internationally recognised standards and risk management frameworks such as AI RMF 1.0 \citep{tabassi_ai_2023},  together with established cybersecurity posture-management practices \citep{cheruku_ai-driven_2025,goodloe_assuring_2022,google_secure_2023, chang_cisco_2025}, were used to define the gating levels of the ladder.

Across each pillar, the ladder defined an ordinal progression (Levels~0--5) that separated static, document-bound assurance from posture-ready assurance:
\begin{itemize}
    \item Level~0 (Ad-hoc): Processes are not defined; assurance is implied rather than specified.
    \item Level~1 (Document-Centric): There are principles or checklists, but the evidence remains static (e.g., PDFs and manual attestations).
    \item Level~2 (Structured/ aware of controls): Risks, controls, roles, or stages are defined, but evidence collection and verification remain largely manual.
    \item Level~3 (Machine-Ingestible): Schemas, APIs, or machine-readable interfaces are specified, allowing partial automation of evidence capture or evaluation.
    \item Level~4 (Posture-Ready): Reusable governance profiles, continuous evidence pipelines, and executable verification logic are specified end to-end to support continuous assurance.
    \item Level~5 (Ecosystem/ Agencies): Posture profiles and signals are shareable and interoperable between organisations, and audit execution can be delegated to autonomous agents.
\end{itemize}

\begin{table}[H]
\centering
\caption{Logic and decision points of the evidence-gated G--O--A Capability Ladder.}
\label{ladder-categories}
\small
\renewcommand{\arraystretch}{6}
\rotatebox{90}{
\begin{tabular}{@{} p{0.14\linewidth} p{0.4\linewidth} p{0.4\linewidth} @{}}
\toprule
\textbf{Pillar} & \textbf{Core evaluation logic} & \textbf{Key decision point (static vs.\ posture-ready)} \\
\midrule
\textbf{G (AI GRC)} & 
Evaluated whether accountable claims were decomposed into auditable objectives, risks, controls, and explicit success thresholds suitable for verification. & 
Level~4 required a transition from narrative checklists to reusable, versionable profiles that encoded claims, control intent, and acceptance thresholds derived from risk appetite. \\ 
\addlinespace
\textbf{O (AI Ops)} & 
Evaluated the form, timeliness, and machine-tractability of evidence generated by operational controls (i.e., what could be collected by default, and at what speed). & 
Level~4 required a transition from episodic document evidence packs to continuous, machine-ingestible telemetry pipelines that could sustain high-velocity assurance. \\ 
\addlinespace
\textbf{A (AI Audit)} & 
Evaluated whether the audit method was specified as executable assurance logic and whether the output was a narrative report or a quantitative trust signal. & 
Level~4 required computational roll-up semantics that aggregated evidence against defined thresholds to produce a continuously updated posture signal. \\ 
\bottomrule
\end{tabular}
}
\end{table}
\FloatBarrier

The ladder was designed to be progressive and non-compensatory. The high capability in one pillar could not compensate for deficits in another because continuous auditing required both machine-ingestible evidence feeds (Operations) and explicit success thresholds (Governance) to support executable roll-up semantics (Audit). Table \ref{ladder-categories} summarises the evaluation logic for each pillar and the decision points that defines Level~4.

The rubric applied an "evidence-first" rubric where capability levels were awarded only if a framework provided explicit verifiable artefacts. Advisory statements or general guidance were insufficient. For example, recommending that an organisation maintain a risk register did not meet the same threshold as defining its required structure, schema, or executable logic.

Figure \ref{fig:Figure4} demonstrates the benchmark results based on Table \ref{tab:table2} in Appendix \ref{sec:apdx:b:frameworks}:
\begin{figure}[H]
    \centering
    \includegraphics[width=1.5\linewidth, angle=90]{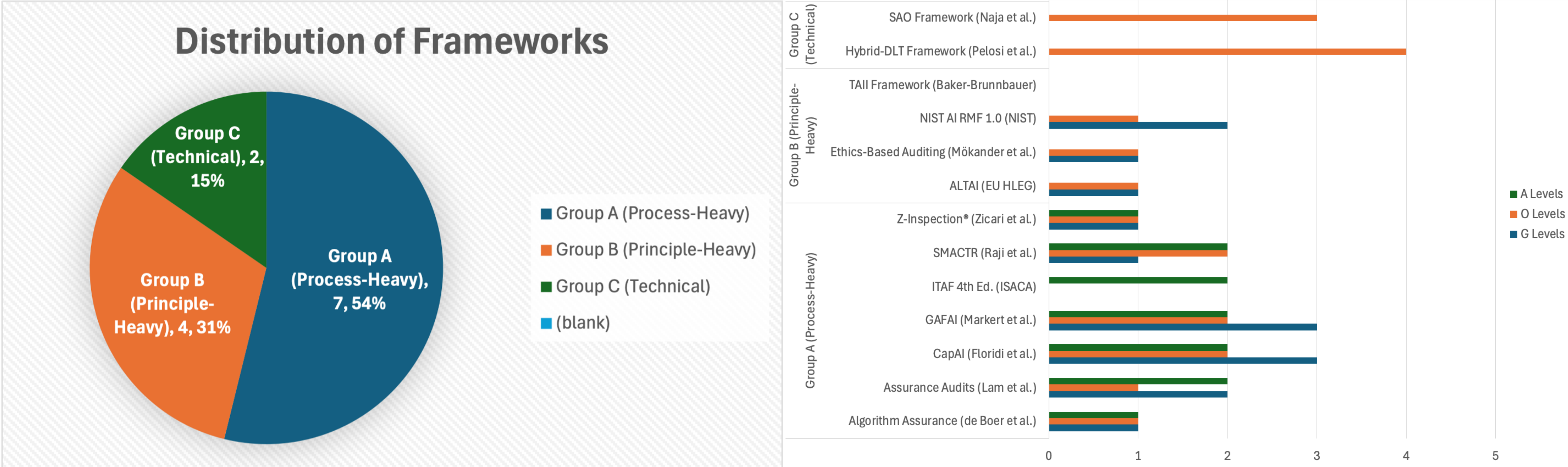}
    \caption{Framework classification distribution results from the capability ladder (pillar scores and morphology groups).}
    \label{fig:Figure4}
\end{figure}
\FloatBarrier

The evaluation of the thirteen frameworks revealed a distinct morphology in the Trustworthy AI audit and assurnace framework corpus, clustering the state of the art into three groups (see Figure \ref{fig:Figure4}) all three groups demonstrate gap between towards achieving continuous Trustworthy AI Assurance at scale.

\paragraph{Group A: Process-Heavy and Document-Centric Frameworks}
  The first group, designated as Group A, includes frameworks such as SMACTR \citep{raji_closing_2020}, CapAI \citep{floridi_capai_2022}, and GAFAI \citep{markert_gafai_2022}. These frameworks show relative maturity in the Audit pillar. They typically achieve Level 2, defined as a repeatable audit workflow. This is because they clearly define audit stages, roles, and required artefacts. For example, SMACTR specifies a five-stage lifecycle from scoping to reflection. CapAI provides a conformity assessment protocol aligned with the EU AI Act. However, their limitations appear in the Operations pillar. Their evidence collection processes are primarily based on document-based artefacts. As a result, they scored at Level 1 (Document evidence pack) or Level 2 (Control-aware evidence collection) for Operations. Because they depend on static artefacts such as datasheets, PDFs, and manual reports, rather than machine-ingestible telemetry, they cannot support high-velocity, continuous trust-signal generation in Agentic AI environments.

\paragraph{Group B: Principle-Heavy and Execution-Light Frameworks}
  The second group, Group B, includes widely recognised policy instruments such as the NIST AI Risk Management Framework \citep{national_institute_of_standards_and_technology_artificial_2023} and the EU Assessment List for Trustworthy AI (ALTAI) \citep{european_commission_white_2020}. These frameworks are strong in the Governance pillar. They often reach Level 2 (Risk-structured governance) and in some cases Level 3. This is because they provide comprehensive taxonomies of ethical principles and structured risk categories. However, the analysis identified a critical limitation in execution logic. These frameworks operate primarily as reflective guidance or voluntary checklists. They do not define an executable audit method. As a result, they receive an Audit score of Level 0 (Audit implied). For example, the NIST AI RMF provides a robust vocabulary for mapping and measuring risk. However, it does not specify the executable assurance logic required to translate these measurements into a formal audit determination. Therefore, it functions as a governance reference framework rather than an operational assurance mechanism.

\paragraph{Group C: Technical and Governance-Disconnected Frameworks}
  The third group, Group C, includes technical proposals such as the Hybrid-DLT Framework \citep{pelosi_hybrid-dlt_2023} and the SAO Framework \citep{verborgh_semantic_2021}. These frameworks show strong capability in the Operations pillar. They typically achieve Level 3 (Machine-ingestible interfaces) or Level 4 (Continuous evidence pipelines). This is because they use technologies such as knowledge graphs and blockchain ledgers to capture structured, immutable audit logs. However, this technical strength is not matched by governance capability. These frameworks generally score Level 0 in the Governance pillar. They do not provide reusable posture profiles or defined risk-appetite thresholds to interpret the data they collect.As a result, although they generate continuous technical evidence, they lack a governance layer to determine what constitutes acceptable performance. Without this interpretive layer, the data streams cannot be translated into a validated claim of trustworthiness.

The central finding of this analysis is the existence of a clear “posture-ready vacancy” in the current state of the art. To enable continuous and automatable assurance in the Large-Scale Era \citep{sevilla_compute_2022, mokander_auditing_2024, jaffri_hype_2023}, a framework must simultaneously achieve advanced capability across Governance, Operations, and Audit. It must define reusable risk structures, support continuous ingestion of operational evidence, and provide a mechanism to aggregate that evidence into a validated trust signal. Evidence-gated coding confirmed that no existing framework satisfies all three conditions.
Instead, the landscape is structurally fragmented. Some frameworks are governance-heavy and checklist-driven (Groups A and B). Others are technically continuous but lack governance interpretation (Group C). None integrate both dimensions into a single, closed-loop architecture. This fragmentation mirrors the disconnections observed during the development of the Trustworthy AI Assurance Ontology. The required elements exist in isolation, but they are not connected. Without integration across governance intent, operational telemetry, and audit validation, a continuous state of trustworthiness cannot be produced. This gap validates the need for a solution to bridge the divide.

\subsection{Conclusion and future research requirements}

The current state of Trustworthy AI audit and assurance frameworks was evaluated, and their readiness to support scalable assurance in high-velocity environments comparable to cybersecurity posture-management systems was assessed. Using an ontology-driven capability ladder and evidence-based coding, the review showed that existing frameworks predominantly supported document- and checklist-based assurance and only partially specified the operational interfaces and decision logic required for continuous monitoring. No framework in the corpus provided an end-to-end, posture-ready configuration.

A posture-based approach reframes the assurance from a point-in-time attestation to a continuously generated trust state that can be observes a signal over time. Conceptually, this requires a repeatable flow in which controls are tested and measured (Test), evidence is interpreted against defined criteria (Evaluate), results are checked against explicit thresholds and risk appetite (Verify), and results are monitored over time to demonstrate sustained performance (Validate) \citep{tabassi_ai_2023}. Realising this flow at scale required reusable governance profiles, machine-ingestible evidence, and continuous audit roll-up semantics that the literature did not provide end-to-end.

These findings indicated a need for further research to design a posture-ready solution to the identified capability gap:
\label{content:RQ1}

\begin{itemize}
    \item [RQ1] What are the essential components, flows, and processes that constitute a Trustworthy AI Posture (TAIP) capable of addressing the vertical and horizontal scale?
    \begin{itemize}
        \item [RQ1.1] To what extent do existing AI assurance frameworks support the transition from static reporting to continuous, horizontally scalable posture-based monitoring?
        \item [RQ1.2] How can a posture-based framework be operationalised to map high-level regulatory guardrails to technical operational evidence?
    \end{itemize}
\end{itemize}

Section \ref{sec:engineering-taip} builds on the gap analysis by developing TAIP as an integrated, posture-ready assurance method that unifies G (AI GRC), O (AI Ops), and A (AI Audit) into a repeatable configuration aligned to Level 4. The TAIP solution establishes the foundation for a Level 5 trajectory (ecosystem-scale profiles, connectors, and agentic audit execution), which this paper positions as future work.

\begin{figure}[H]
    \centering
    \includegraphics[width=0.75\linewidth]{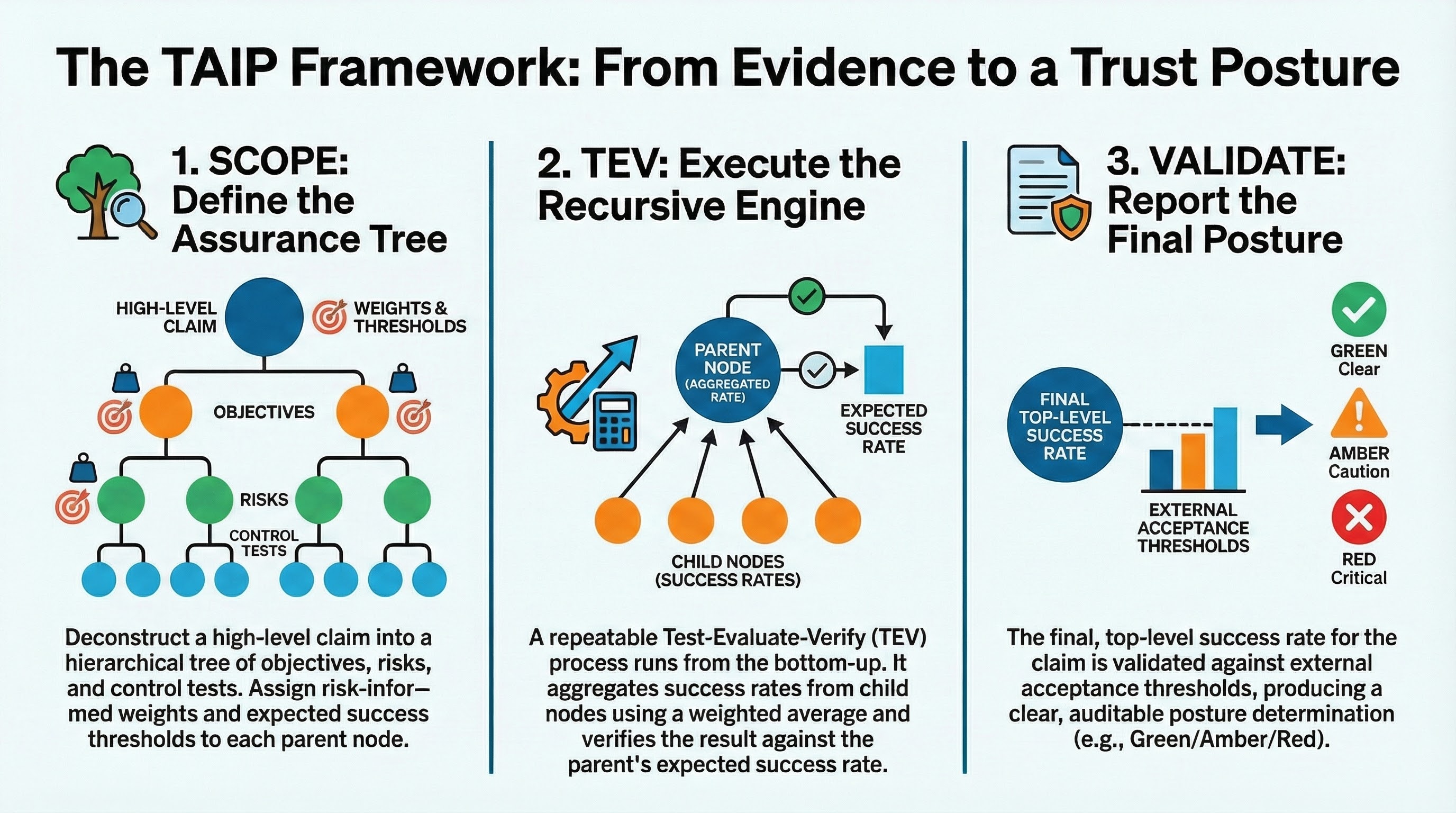}
    \caption{TAIP workflow overview: scope, TEV execution, and validation}
     \label{fig:taip_scope_validate}
\end{figure}
\FloatBarrier

\section{Engineering Trustworthy AI Posture: A Dual-Scale Framework for Agentic AI Continuous Assurance}
\label{sec:engineering-taip}
Based on the assurance gaps and requirements identified in Section \ref{sec:state-of-the-art-lit-review}, the following solution specifies the engineering of the Trustworthy AI Posture (TAIP) framework. TAIP is presented as a conceptual, implementation-agnostic architecture. Although specifically engineered to support the automation of continuous quantified trust signals, its underlying logic remains valid for both manual human-driven audits and fully automated agentic environments. The specification proceeds through six cumulative layers, moving from high-level architecture to validatable application:

\begin{enumerate}
    \item The high-level solution overview of the TAIP framework (\ref{subsec:taip-at-a-glance}).
    \item The research foundations and design methodology, covering the principles, constraints, and convergence strategy involved in the creation of TAIP (\ref{subsec:taip-research-foundation}).
    \item TAIP conceptual engineering, defining the core information structures and the functional design of the assurance engine (\ref{subssec:taip-constructs}).
    \item The practitioner's playbook, describing the canonical workflow for implementing the framework, illustrated via a running example (\ref{subsec:taip-playbook}).
    \item The application use case, which demonstrates practical utility by operationalising the Australian AI Guardrails (Guardrail No. 3) within a Microsoft 365 Copilot deployment (\ref{subsec:taip-co-pilot-example}).
    \item The validation of the TAIP solution, the qualification of design principles against the results of the use case, and established benchmarks (\ref{subsec:taip-validation}).
\end{enumerate}

\subsection{TAIP "at a glance": Trustworthy AI Posture as an Assurance Signal} \label{subsec:taip-at-a-glance}

TAIP framework is a purpose-built Trustworthy AI Assurance method designed to produce a native output that is both human-interpretable and machine-actionable: \textbf{a quantified posture signal for a specific accountable claim}. 

TAIP standardises the method by which trust claims are tested, evaluated, verified, and validated. It operationalises the TEVV logic of the NIST AI RMF 1.0 \citep{tabassi_ai_2023} as an executable assurance pattern.
The framework defines how trust signals are generated from management claims to risks, controls, and operational evidence while allowing policy content to vary \citep{brundage_toward_2020}. By separating the checklist (what must be satisfied) from the assurance method (how it is tested and monitored), TAIP makes it possible to model policy requirements and assurance mechanics independently. Therefore, multiple checklists can be applied using the same signal-generation process. This architectural decoupling is the primary enabler of scalability \citep{ojewale_towards_2024}.

To achieve this, the framework is designed as a deterministic Input–Process–Output (IPO) workflow, as illustrated in Figure \ref{fig:taip_scope_validate}. The cycle covers three well defined steps :

\paragraph{Step 1 (Input-Scope)}
\label{taip-at-glance-input-scope}
Assurance begins with the definition of scope through a structured Posture Profile. This is where the auditor records the accountable management claims and regulatory requirements of the organisation. These high-level claims are decomposed into specific objectives using a decomposition matrix. The AI System Components / Life Cycle Decomposition Matrix (Table \ref{Decomposition AI System Matrix}) forces abstract claims into the concrete Operational Design Domain (ODD) \citep{herrera-poyatos_responsible_2025}. Then it classifies them across relevant stages of the life cycle \citep{agarwal_seven-layer_2024,isoiec_iso_2022,iso_iso_2022-1,isoiec_iso_2022-1}.
At this stage, risks and mitigation objectives are identified for each claim. Each objective is explicitly bound to the sources of operational evidence required for the audit. This is where the governance intent is connected to measurable system artefacts.

\paragraph{Step 2 (Process-TEV)}
This step executes the audit sequence. It performs Test, Evaluate, and Verify (TEV) based on the AI RMF 1.0 methodology \citep{tabassi_ai_2023}. Operational tools generate normalised evidence records. The TEV engine ingests these records and computes the performance outcomes for each node in the posture tree. Each result is compared with predefined, risk-informed thresholds assigned to the objectives.
This stage represents the active audit process.

\paragraph{Step 3 (Output-Validate)}
The final step performs validation. The results are assessed against the expected success criteria defined during the scoping. This produces a definitive posture determination and a traceable chain of evidence. In traditional terms, this is the equivalent of producing an audit report.
Each execution generates a bounded “atomic” trust signal. Continuity is achieved by repeating this process on a schedule or through event-driven triggers. Over time, this creates a time series of posture signals for the same claim.

\begin{figure}[H]
    \centering
    \label{HORIZON-OF-TRUST}
    \includegraphics[width=1\linewidth]{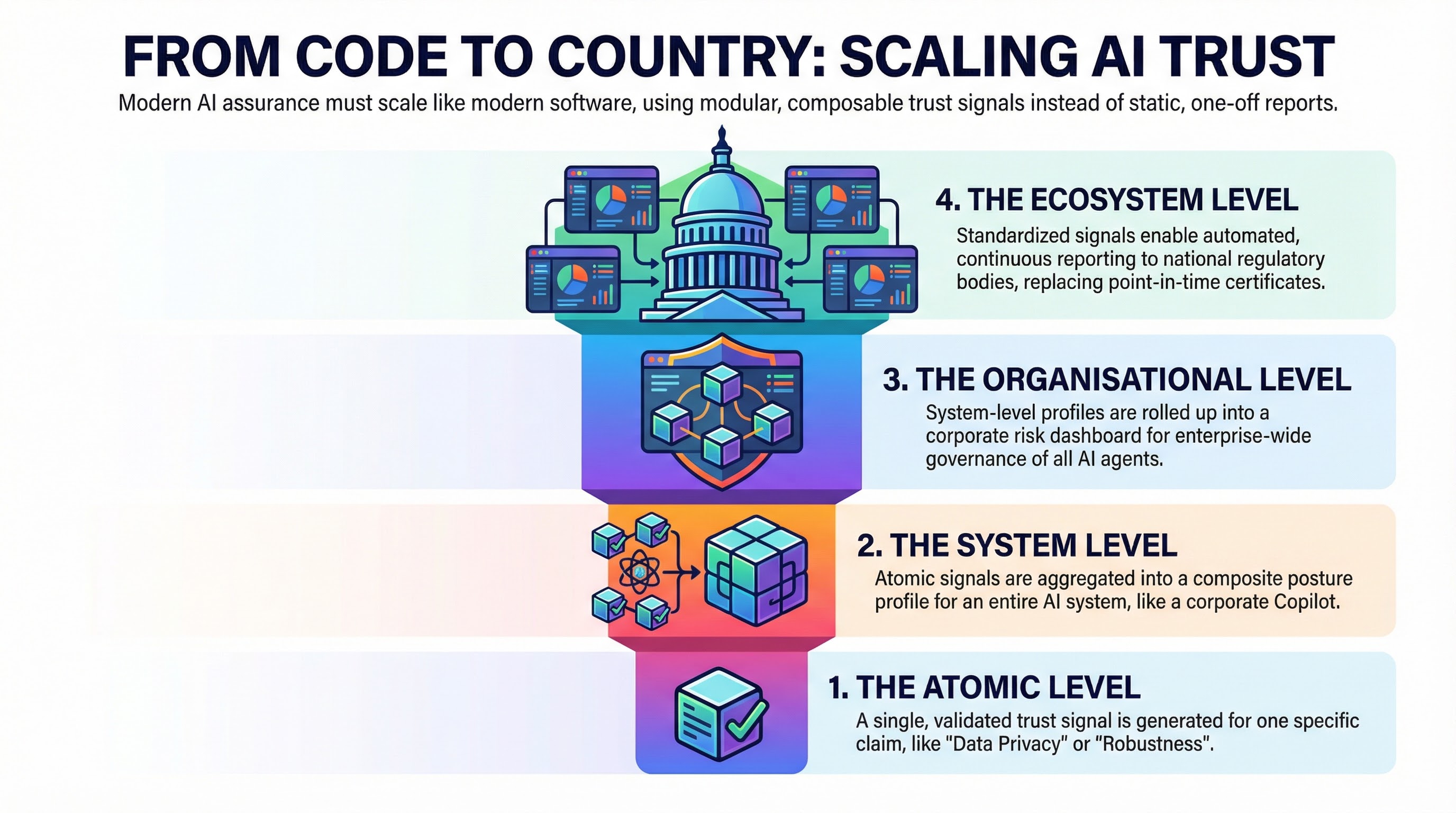}
    \caption{Vertical scaling of posture signals from atomic to ecosystem level - the "horizon" of Trustworthiness }
    \label{fig:Figure6}
\end{figure}
\FloatBarrier

The TAIP workflow follows a simple Input–Process–Output cycle. Each execution Test, Evaluate, and Verify one specific management claim and produces one validated trust signal. This signal represents an atomic unit of trust.
As shown in Figure \ref{fig:Figure6}, the framework is designed to scale by composition. 
\paragraph{Level 1 (Atomic)}
TAIP evaluates a single claim, such as privacy or robustness. The result is a clear pass or fail posture signal for that claim \citep{naja_using_2022}.
\paragraph{Level 2 (System)}
, multiple atomic signals are combined to form a system posture profile. A single AI system may satisfy some claims and fail others. The overall system posture reflects this combination.
\paragraph{Level 3 (Organisation)}
, posture profiles from many systems are aggregated. An enterprise that operates hundreds of AI agents can monitor its overall risk position in real time.
\paragraph{Level 4 (Ecosystem)}
, validated posture signals can be shared beyond organisational boundaries. This enables regulators, industry bodies, or government schemes to continuously monitor trust rather than relying on static certificates \citep{javadi_monitoring_2021,naip_common_2020}.

This model mirrors the evolution of cybersecurity assurance \citep{european_union_agency_for_cybersecurity_ai_2020}. In cybersecurity, Common Criteria established repeatable methods for evaluating and recognising security claims across jurisdictions \citep{naip_common_2020}. TAIP provides the structural foundation for the trust eco-system that supports the original ambition of the European Commission and the EU AIA \citep{european_commission_white_2020,noauthor_regulation_2024}. 

TAIP’s novelty is a method-level shift: assurance is engineered as an \emph{executable object}, not a static report.
Traditional audit frameworks fuse the checklist (what must be satisfied) with the audit procedure (how it is executed), so changing requirements forces the workflow to change\citep{mokander_auditing_2024,minkkinen_continuous_2022,li_making_2024,li_trustworthy_2023}.
TAIP applies an object-oriented separation of concerns\citep{gamma_elements_1995}.
Policy content, risks, and control sets remain configurable, while the TEVV execution semantics remain stable and reusable\citep{national_institute_of_standards_and_technology_artificial_2023,tabassi_ai_2023}.
This decoupling removes a primary scalability constraint and enables the same assurance mechanics to operate across multiple regulatory environments.

This is critical for agentic systems.
Agents can be ephemeral, replicated, and non-deterministic.
TAIP treats assurance as a runtime primitive that can be instantiated with an agent, scoped to its specific run, and executed at machine speed.
Assurance therefore becomes continuous and compositional, even when an agent’s lifetime is measured in seconds or milliseconds.

TAIP also closes the governance-to-operations gap by binding intent to evidence\citep{mokander_us_2022,verborgh_semantic_2021,brundage_toward_2020}.
Claims are decomposed into objectives and linked directly to operational evidence sources.
Evidence is then aggregated into a single posture outcome (success/fail)\citep{barclay_providing_2023}. \label{content:supermarket-bill}
This is analogous to a supermarket bill: many different items are reduced to one total through a common unit of account.

By standardising execution while allowing policy to vary, TAIP enables vertical traceability and horizontal automation for trustworthy assurance at scale.


By standardising the audit execution method while allowing policy content to vary, TAIP addresses both dimensions of the scalability challenge. Vertically, it supports diverse governance and control environments and horizontally, it creates a pathway for automation across distributed and Agentic systems.

The result is a Trustworthy AI Posture Assurance object, which can be instantiated at machine speed, across many types of environment, with different policy settings but with a single composable machine ingestible reporting structure, this is how TAIP is operating at the speed and scale of modern AI systems \citep{wu_autogen_2023,yao_react_2022,bloomfield_assessing_2022}. 
The following Section are going to provide the detailed technical specifications behind the engineering and design of TAIP.

\subsection{Research Foundations and Design Methodology} \label{subsec:taip-research-foundation}
TAIP is grounded in recognised risk management standards and implemented through architectural design principles aligned with large-scale, agentic systems. The following sections present its theoretical foundations, define the governing architectural principles, and explain how these converge into the AI Assurance Object as the operational mechanism for generating trust signals.

\subsubsection{Theoretical Foundations}
TAIP constructs its assurance logic by synthesising three established domains. This ensures that while the application (AI agents) is novel, the assurance mechanics remain rooted in defensible audit practices.

\paragraph{Risk Management (The "Why")}: TAIP adopts the NIST AI Risk Management Framework (AI RMF 1.0)\citep{tabassi_ai_2023}, specifically its TEVV (Test, Evaluate, Verify, Validate). Although NIST provides the \textit{vocabulary} for assurance, it does not prescribe the \textit{mechanism} for continuous machine-speed auditing. TAIP adopts TEVV as its core execution loop.
\paragraph{Audit Defensibility (The "Standard")}: To ensure signals are legally and operationally defensible, TAIP mirrors the logic found in the {COSO Internal Control framework} \citep{calagna_applying_2021}. This requires that a control be not just "checked" but also evaluated for design adequacy, operating effectiveness, and sustainability.
\paragraph{Systems Thinking (The "Scope")}: Assurance is treated as a structural property of the system \citep{willett_systems_2022}. This involves decomposing the AI system into boundaries (Operational Design Domains)\citep{herrera-poyatos_responsible_2025} to ensure that evidence is bound to specific AI System components rather than vague generalities.

\subsubsection{Architectural Principles and Constraints}
To address the "Agentic AI Assurance Scalability Crisis" we discuss in Section \ref{sec:intro} characterised by the rapid proliferation of ephemeral, modular agents, TAIP is governed by three core design principles. These principles standardise the "grammar" of assurance while allowing the "vocabulary" of risks to remain context-dependent.

\subparagraph{The Policy Abstraction Pattern (Managing Divergence)}
A fundamental principle is the variability of policy across jurisdictions \citep{adedokun_global_2024, mokander_ethics-based_2021-1} (e.g., EU AI Act \citep{noauthor_regulation_2024} vs. Australian Guardrails \citep{government_safe_2023}). TAIP decouples regulatory obligations/organisational policies from technical implementations \citep{chen_security_2024}. The framework ingests policy as a hierarchy of \textit{Claims} and \textit{Objectives}, abstracting the "intent" of the regulation from the "mechanics" of the test. This ensures that the architecture remains resilient to changing regulatory landscapes without requiring code-level refactoring.

\subparagraph{The Normalized Evidence Stream (Managing Heterogeneity)}
The evidence in AI systems is inherently heterogeneous, spanning automated logs, red-teaming reports, and static analysis \citep{minkkinen_continuous_2022}. TAIP employs a \textit{normalization principle} in which the "decisioning logic" of a test is external to the framework. Evidence is ingested as a standardised binary schema (Success/Fail). This abstraction ensures that the assurance engine focuses on \textit{aggregation} and \textit{binding} rather than the subjectivity of specific testing tools \citep{mittelstadt_principles_2019}.

\subparagraph{The Recursive Text Evalute Verify (TEV) Pattern (Managing Scale)}
To scale horizontally across distributed agents and vertically from technical controls to governance claims, the TEV process is designed as a \textit{Recursive Pattern} \citep{gamma_elements_1995}. The assurance logic is a "frozen," reusable asset applied identically at every layer of the system hierarchy. A "posture" is not a static report, but a dynamic instance created by recursively aggregating signals from the bottom up.

\subsubsection{The Convergence Strategy: The Trustworthy AI Assurance Object}
\begin{figure}[H]
    \centering
    \includegraphics[width=0.5\linewidth]{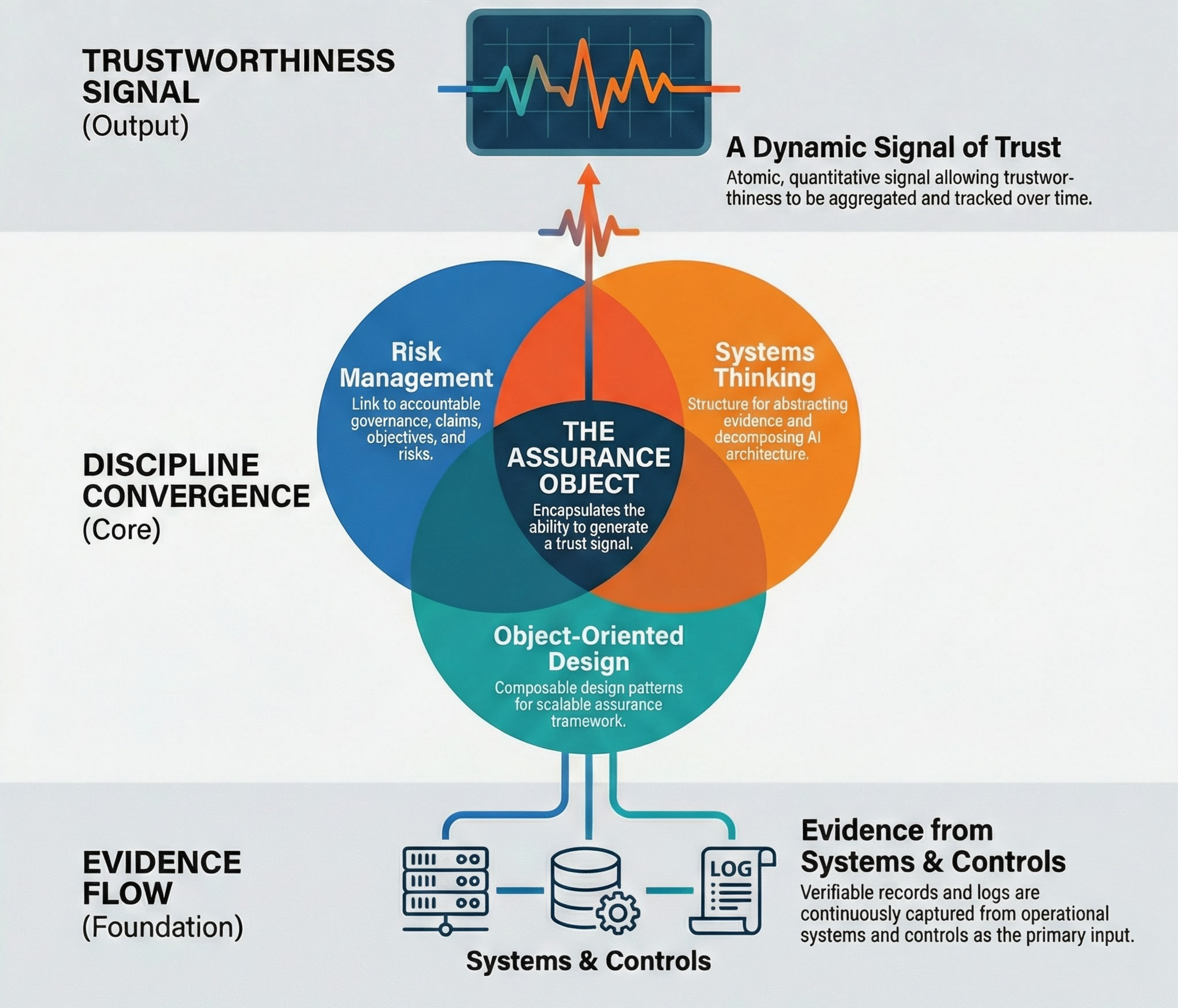}
    \caption{Convergence of three practices to transform TEVV audit practice into a Signal}
    \label{fig:TAIP-VENN}
\end{figure}

To operationalise these principles, TAIP uses a "Convergence Strategy" (see Figure \ref{fig:TAIP-VENN}). This strategy refactors the abstract (Test Evaluate Verify) and Validate lifecycle into a concrete executable software construct: the \textbf{AI Assurance Object}.
\FloatBarrier

\begin{figure}[H]
    \centering
    \includegraphics[width=0.75\linewidth]{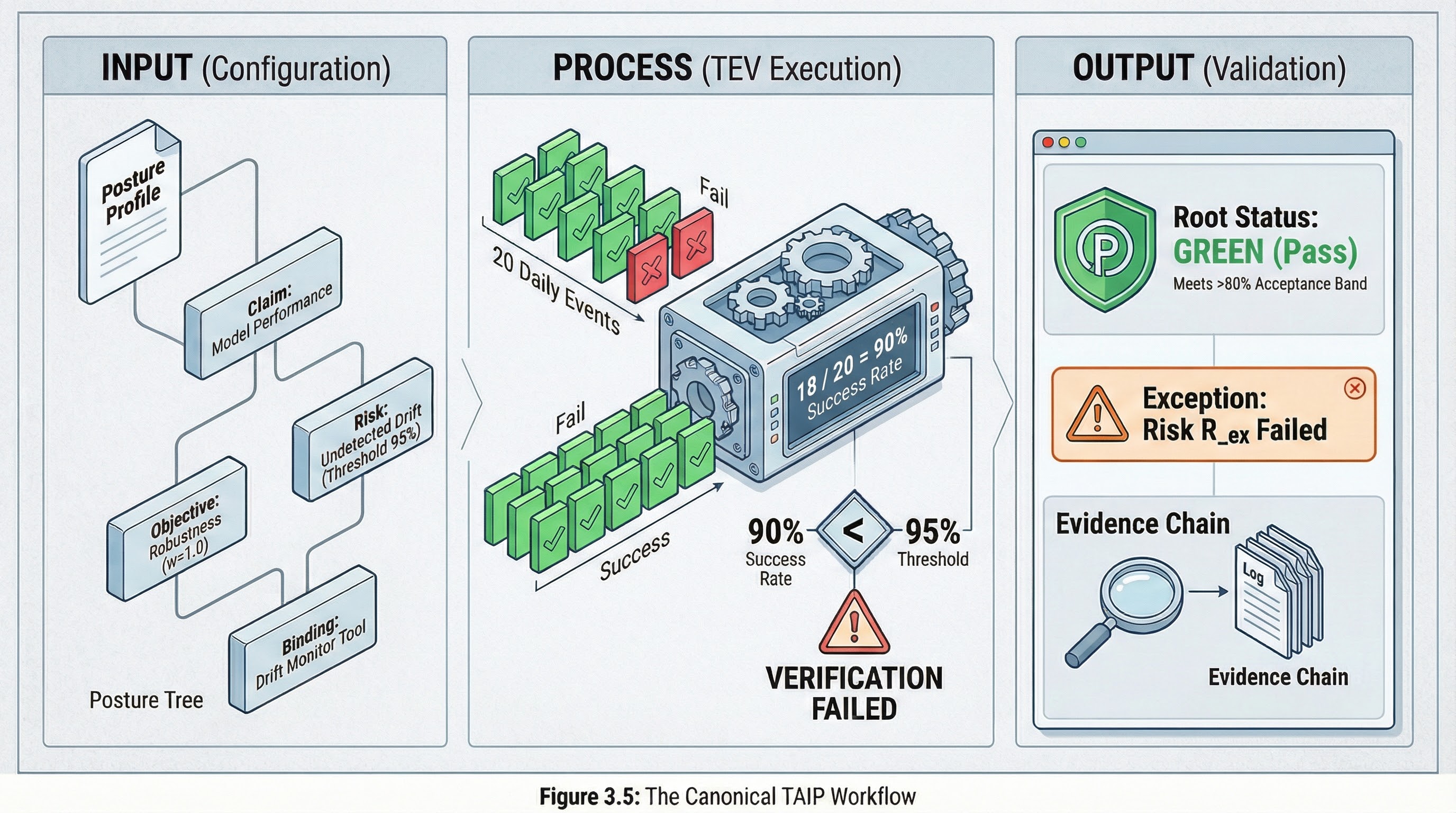}
    \caption{AI Assurance Object Input-Process-Output}
    \label{AI Assurance Object IPO}
\end{figure}

The AI Assurance Object is the foundation of the TAIP framework. It encapsulates the Input–Process–Output (IPO) workflow required to generate a trust signal (\ref{AI Assurance Object IPO}):

\begin{itemize}
    \item \textbf{Input (The Binding):} Derived from GRC practices, this phase links an organisational \textit{Claim} to a specific AI system component using a decomposition matrix (Table \ref{Decomposition AI System Matrix}). This ensures every signal is "bound" to a specific context (e.g., "Training Data" vs. "Runtime Model").
    \item \textbf{Process (The TEV Loop):} Derived from Object-Oriented Design, this is the runtime execution of the Test-Evaluate-Verify cycle. This process is applied recursively across the posture tree.
    \item \textbf{Output (The Signal - Validate):} Derived from Audit Logic, this validates the result against a threshold to issue a quantifiable posture determination (Satisfied/Not Satisfied).
\end{itemize}

\FloatBarrier

\begin{table}[H]
  \centering
  \resizebox{\linewidth}{!}{%
  \begin{tabular}{@{}lcccccc@{}}
    \toprule
     & IDEA & DESIGN & DEVELOPMENT & DEPLOY & OPERATE & RETIRE \\
    \midrule
    IP &  &  &  &  &  &  \\
    DATA &  & (example) Training data should not be used for production &  &  &  &  \\
    MODEL &  & &  &  & (example) Model should be monitored hourly for drift &  \\
    ALGORITHM &  &  &  &  &  &  \\
    INFRASTRUCTURE &  &  &  &  &  &  \\
    CLOUD SERVICES &  &  &  & (example) Data Residency should be applied based on Geography &  &  \\
    HCI &  &  &  &  &  &  \\
    COMMERCIAL &  &  &  &  &  &  \\
    ACTORS &  &  &  &  &  &  \\
    \bottomrule
  \end{tabular}%
  }
  \caption{AI System/Lifecycle Decomposition Matrix with examples}
  \label{Decomposition AI System Matrix}
\end{table}
\FloatBarrier

By encapsulating the TEV-V cycle within this standardized object, TAIP solves the "chain of evidence" problem \citep{pelosi_hybrid-dlt_2023}. It creates a traceable path from the lowest technical log (Input) to the highest governance claim (Output), enabling the "human-interpretable, machine-actionable" signal defined in the introduction.

\subsection{TAIP constructs and operational logic} \label{subssec:taip-constructs}

TAIP can operate in either manual or automated environments. The underlying method remains the same in both cases.
The framework follows a standard Input–Process–Output (IPO) structure. Governance claims are captured as structured inputs. These claims are then processed through a defined assurance sequence. The result is a clear, validated trust signal. Whether executed by humans or by software, the logic does not change. High-level accountability claims are systematically decomposed into measurable objectives and bound to operational evidence\citep{tabassi_ai_2023,minkkinen_continuous_2022,calagna_applying_2021}.

\subsubsection{The Input Constructs}
\paragraph{The Trustworthy AI Assurance Unit (TAI-AU)}

The foundational building block of the framework is the basic TAIP atomic unit, it represents the smallest atomic encapsulation of governance intent and technical reality. Every execution of TAIP, regardless of complexity, is composed of these units.

\begin{figure}[H]
    \centering
    \includegraphics[width=1\linewidth]{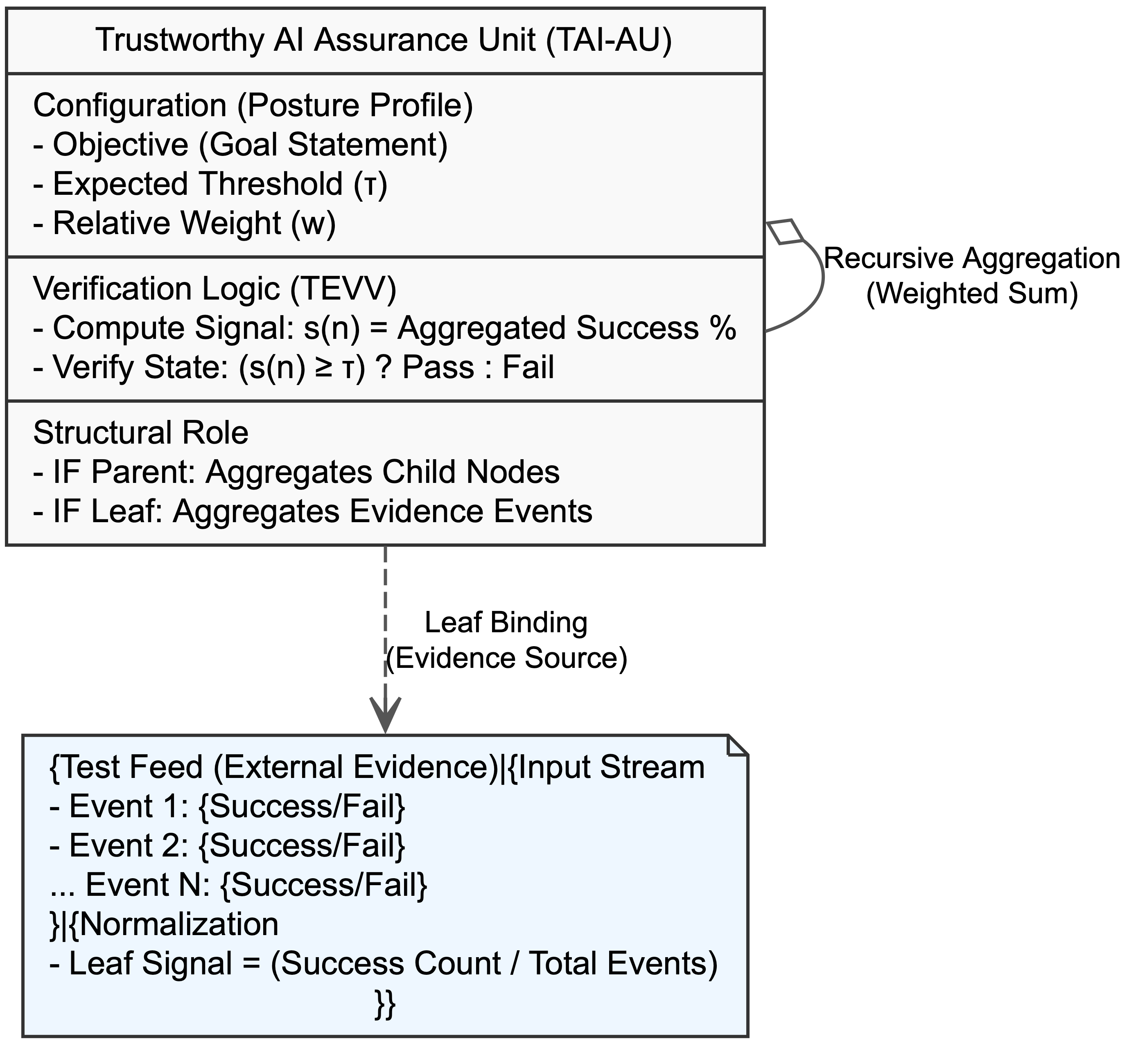}
    \caption{TAIP Atomic Unit (node) }
    \label{fig:taip-node-tai-au}
\end{figure}

A TAI-AU consists of three tightly coupled variables \ref{fig:taip-node-tai-au}:
\begin{itemize}
    \item \textbf{The Objective:} A specific goal derived directly from the \textbf{Decomposition Matrix}. This ensures that every objective is bound to a specific AI system component (e.g., training data, model weights, or inference API) within its specific lifecycle stage\citep{herrera-poyatos_responsible_2025,isoiec_iso_2022,agarwal_seven-layer_2024,national_institute_of_standards_and_technology_artificial_2023}.
    \item \textbf{The Risk(s):} One or many identified threats that could lead to the failure of the objective. This is the technical binding layer where the threat model meets the system architecture.
    \item \textbf{The Test Binding:} Each risk is bound 1:1 to a specific test feed\citep{minkkinen_continuous_2022,toader_auditability_2019,pelosi_hybrid-dlt_2023}. This feed is an external signal sourced from specialized AI testing tools that provides a binary \textbf{Success/Fail} outcome.
\end{itemize}
To make this unit evaluative, the user manually defines an \textbf{Expected Success Rate} for each risk\citep{calagna_applying_2021,isaca_cobit_2018}. This rate represents the organization’s risk appetite stating, for example, that the risk of model hallucination must be successfully mitigated in 98\% of observed events to satisfy the objective.

\paragraph{The Posture Tree: Recursive Composition}
While the TAI-AU is the atomic base, AI systems require a multi-faceted approach to trust. The \textbf{Posture Tree} is the structural mechanism that allows these atomic units to be organized into a comprehensive hierarchy through the capability of \textbf{Recursive Aggregation} as an attribute of the Composite Design\citep{gamma_elements_1995} we discussed at TAIP at a glance\ref{content:supermarket-bill}. 
\begin{figure}[H]
    \centering
    \includegraphics[width=0.5\linewidth]{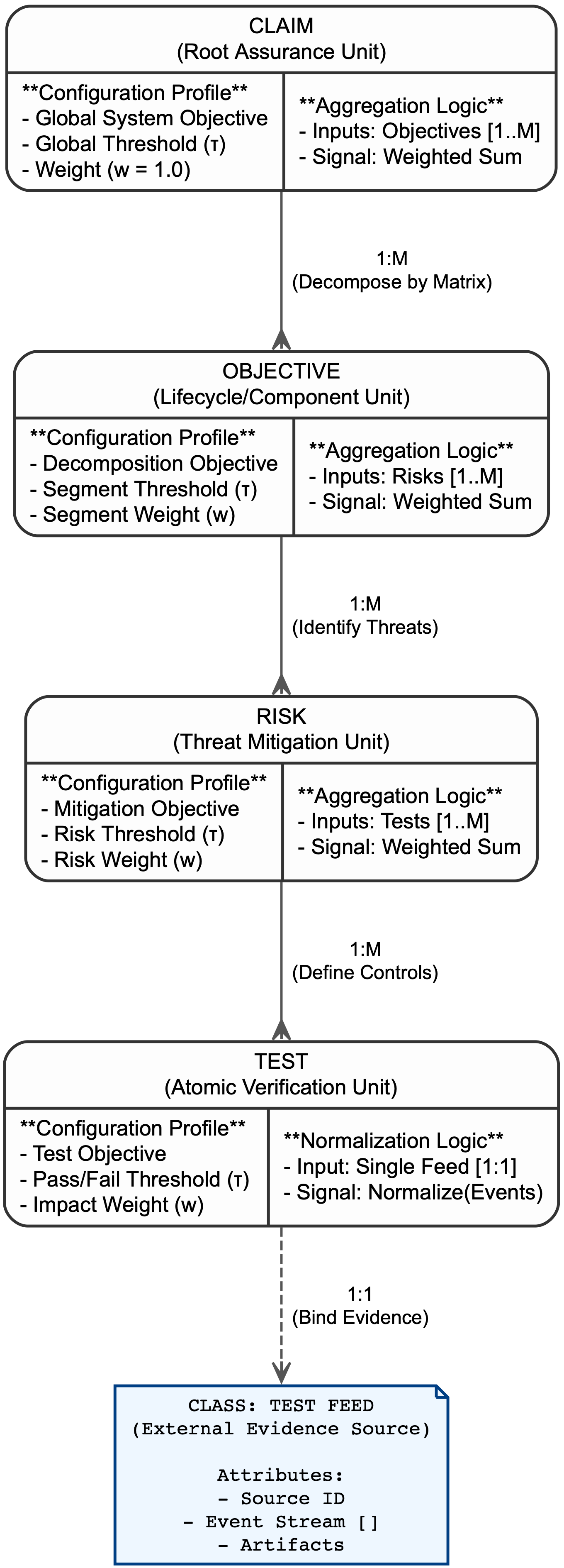}
    \caption{TAIP Posture Tree}
    \label{fig:taip-recursive-tree}
\end{figure}
By nesting TAI-AUs within one another, the framework allows for arbitrary depth \ref{fig:taip-recursive-tree}: 

\begin{itemize}
    \item \textbf{The Leaf Level:} The technical termination points where specific risks meet their external test feeds.
    \item \textbf{The Node Level:} Aggregation points representing a collection of risks, a component in its lifecycle stage. Each node aggregates the success/fail signals from its children, Each node has an expected minimal success rate threshold
    \item \textbf{The Root Level (The Claim):} In this model, the \textbf{Accountability Claim} is simply the primary, top-level objective of the tree.
\end{itemize}

\paragraph{Structural Manifestation and AI Lifecycle Binding}
\begin{figure}
    \centering
    \includegraphics[width=0.75\linewidth]{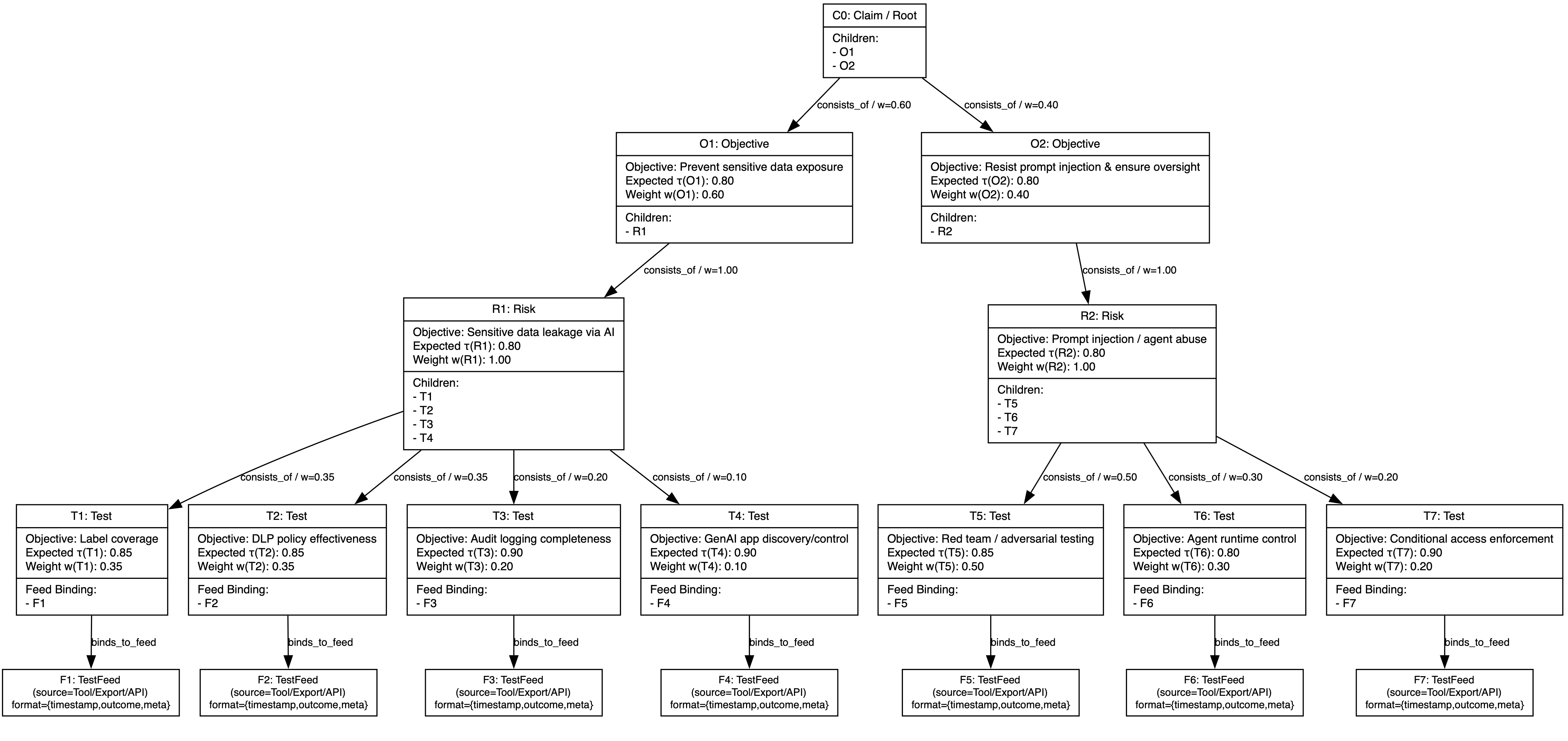}
    \caption{TAIP Complex Manifestation}
    \label{fig:taip-complex-tree}
\end{figure}
The use of the Posture Tree ensures that the evaluation logic remains identical whether the tree is simple or highly complex. The depth of the tree is determined by the complexity of the AI system's lifecycle as defined in the decomposition matrix \ref{Decomposition AI System Matrix} \citep{agarwal_seven-layer_2024,isoiec_iso_2022,agarwal_seven-layer_2024}.

\begin{enumerate}
    \item \textbf{Simple Manifestation:} A "shallow" composite tree where a single Claim (as the root objective) points directly to a few Risks and their corresponding Tests.
    \item \textbf{Complex Manifestation:} A "deep" composite tree where a Claim decomposes into multiple sub-objectives. Each sub-objective represents a unique intersection of an AI component and a lifecycle stage (e.g., "Data Privacy during Collection" vs. "Model Robustness during Deployment") \ref{fig:taip-complex-tree}.

\end{enumerate}

Using this recursive information structure, TAIP transforms fragmented manual oversight into a unified, composable assurance architecture. Each node at every level demands the same manual inputs: \textbf{Weighted Importance} (relative priority) and \textbf{Success Criteria} (policy-driven thresholds), creating a standardised schema for continuous, lifecycle-aware AI assurance.

\subsubsection{The Process constructs: the posture tree and composite aggregation}
Once the Posture Profile is defined, execution occurs within the AI-System Posture Tree. The tree is the information structure of the audit and captures an organised set of claims, objectives, risks, and evidence in a hierarchy that mirrors the architecture of the AI system.
The process at each node follows a simple recursive TEV sequence\citep{tabassi_ai_2023,gamma_elements_1995,fan_decentralized_2021,bloomfield_assessing_2022}.
\paragraph{Test}:
The node harvests results from its bound evidence sources or child nodes. These results represent measured control performance.
\paragraph{Evaluate}:
The node calculates a weighted score. Each child result is multiplied by its assigned risk weight. The combined value reflects organisational risk appetite defined during the scoping phase.
\paragraph{Verify}:
The calculated score is compared against the predefined threshold for that node. If the score meets or exceeds the expected success rate, the node is marked as satisfied. If it falls below the threshold, the node fails.

Once a node posture (success or failure) is determined, it becomes an input to its parent. The same sequence   Test, Evaluate, Verify   is applied again at the next level. In this way, posture is calculated recursively from the lowest technical controls up to the highest governance claim.
The result is a compositional trust signal. Each node produces a clear outcome. These outcomes scale upward without manual interpretation, allowing assurance to operate consistently across complex, distributed systems.

\subsubsection{The Output constructs: validation and posture determination}

The final stage of the workflow is Validation. This step determines the overall status of the assurance claim.
Earlier steps verify individual components against technical thresholds. Validation is different. It assesses the overall posture against the organisation’s defined risk appetite \citep{calagna_applying_2021}.
The root node produces a composite success score. This score is then compared to an organisational threshold. The result is translated into a simple status indicator, such as a Red–Amber–Green (RAG) model.
A Green status means the claim meets the required confidence level.
An Amber status indicates deviation within acceptable risk limits and may require monitoring.
A Red status signals a breach of risk appetite and requires corrective action.
This step converts a technical score into a clear governance decision. It provides accountable management with a human-interpretable signal to support decisions about deploying, continuing, or withdrawing the AI system.

\subsection{TAIP process - how to use the AI Assurance Object} \label{subsec:taip-playbook}

With the architectural foundations established, the next section presents the execution playbook. It follows the defined workflow step by step and uses a running example to demonstrate how a bounded claim produces an atomic trust signal. TAIP can be executed manually today or implemented programmatically without altering its semantics. Each execution produces one validated posture outcome for a defined claim and evaluation window\citep{minkkinen_continuous_2022,tabassi_ai_2023,javadi_monitoring_2021}.
\begin{figure}[H]
    \centering
    \includegraphics[width=1\linewidth]{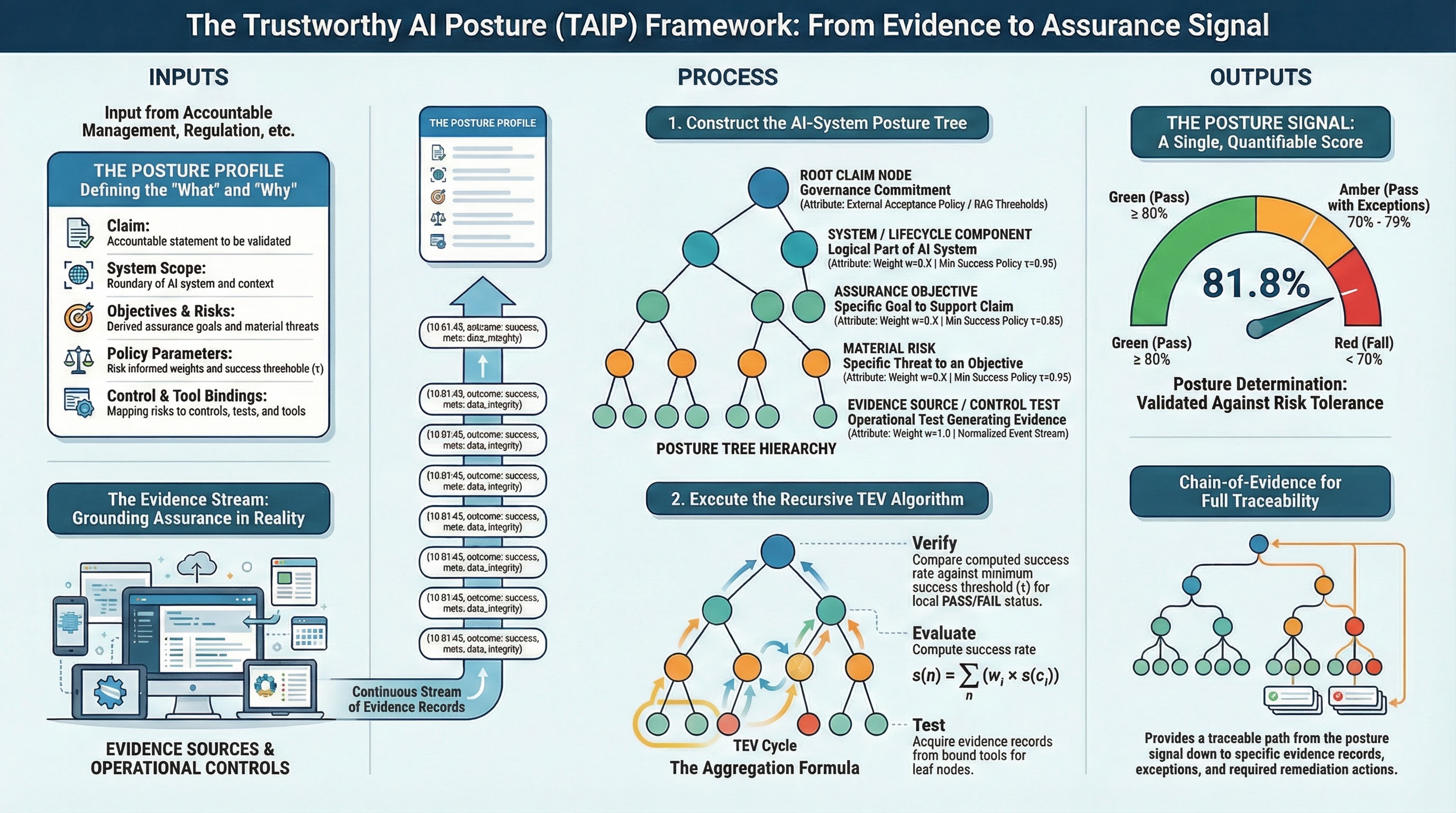}
    \caption{TAIP input-process-output view: posture profile, recursive TEV algorithm, and posture determination.}
    \label{TAIP-INPUT-METHOD}
\end{figure}
\FloatBarrier
\subsubsection{Step 1 (Input): Scope construction and posture profile configuration}

Scope construction is the configuration phase. It happens once per posture run (and then again whenever the posture profile is revised). The output is a configured posture profile and an executable posture tree with: node types and relationships, weights, expected success thresholds, and leaf bindings to tools/feeds that can emit evidence.

\begin{enumerate}
    \item  Claim formalisation (Accountable Management) Formalise the accountable claim that management is willing to stand behind (e.g., a regulatory guardrail, an internal policy commitment, or a contractual requirement). The claim becomes the root node.
    \item  System boundary and decomposition (Operations): Define the in-scope system boundary and decompose the AI system into components and lifecycle stages where controls and evidence exist (e.g., model, retrieval, orchestration, plugins/tools, logging/monitoring, human oversight)\citep{herrera-poyatos_responsible_2025,isoiec_iso_2022,agarwal_seven-layer_2024,national_institute_of_standards_and_technology_artificial_2023}.
    \item   Objective specification and risk identification (GRC): Decompose the claim into AI Assurance Objectives (what must hold true) and associated risks (what could cause failure). Record risks in structured form to support prioritisation and traceability.
    \item  Weighting and expected success thresholds (GRC policy parameters): Assign weights at each 1:M node to represent relative importance. Configure expected success thresholds τ(n) to represent minimum acceptable performance, derived from risk appetite\citep{calagna_applying_2021,isaca_cobit_2018}.
    \item  Evidence binding (Audit–Operations integration) :Bind each risk to one or more control tests and bind each control test to evidence sources (tools, logs, red-team runs, sampling). Ensure each binding yields normalised success/fail evidence records.
\end{enumerate}

\begin{table}
\centering
\begin{tabular}{l}
Running example (used throughout Section 3.4): minimal ``unit of trust'' posture profile \\
\\
1. Claim ($C_{ex}$): The deployed model maintains acceptable performance over time; \\
drift is detected and addressed within policy thresholds. \\
\\
1.1. Objective ($O_{ex}$): Operational robustness and monitoring. \\
\\
1.1.1. Risk ($R_{ex}$): Undetected model drift causes unacceptable degradation. \\
\\
1.1.1.1. Control test ($T_{ex}$): A drift monitoring pipeline runs daily and emits \\
success/fail (within threshold vs breach). \\
\\
Policy parameters (illustrative): \\
$\mathrm{weight}(O_{ex}) = 1.0$ \\
$\mathrm{weight}(R_{ex}) = 1.0$ \\
$\tau(R_{ex}) = 0.95$ \\
root acceptance band: Green $\geq 0.90$ \\
Amber $0.80\text{--}0.90$ \\
Red $<0.80$.
\end{tabular}
\end{table}

\subsubsection{Step 2 (Process): TEV execution and recursive posture aggregation}

Step 2 executes the Test–Evaluate–Verify (TEV) cycle bottom-up across the entire configured posture tree. The defining design choice is recursion: the same TEV pattern is applied at every one-to-many (1:M) node. Adding depth to the hierarchy (e.g., adding a System or Component layer) changes the tree structure, but it does not change the evaluation algorithm\citep{tabassi_ai_2023,gamma_elements_1995}.

\begin{enumerate}
    \item Test (Acquire Evidence). The system acquires evidence records from the bound feeds or tools for the specified evaluation window. To ensure auditability, each record must indicate a binary outcome (success or fail) at a specific timestamp, along with optional metadata\citep{toader_auditability_2019,pelosi_hybrid-dlt_2023}.
    \item Evaluate (Compute Success).The method computes the s(n) [Success Rate of Node n] for every node in the tree.
    \item For a Test Node, the s(T) [Test Success Rate] is calculated as the fraction of success events observed over the total events.
    \item For a Parent Node (such as a Risk or Objective), the success rate is calculated as the weighted aggregation of its children's success rates.
    \item Verify (Compare to Threshold). The method compares the calculated s(n) [Success Rate] against the τ(n) [Expected Success Threshold] defined in the posture profile.
    \item A node is marked PASS if the s(n) [Success Rate] is greater than or equal to the τ(n) [Expected Threshold].
    \item Otherwise, the node is marked FAIL. A failed verification immediately generates an exception record and creates a remediation obligation for that specific branch of the tree.
\end{enumerate}

\begin{table}
\centering

\begin{tabular}{l}
\textbf{\textbf{Running example: TEV computation and local exception (risk fails while root can still validate)}}

Evaluation window: 20 days (20 drift-monitor events).

Observed evidence: 18 success events, 2 fail events.

Compute s(R\_ex)=18/20=0.90.

Verify risk: s(R\_ex)=0.90 < τ(R\_ex)=0.95 → FAIL.

Aggregate objective: s(O\_ex)=s(R\_ex)=0.90 (only one child, w=1.0).

Verify objective: if τ(O\_ex)=0.90 and the convention is s(n) ≥ τ(n), then O\_ex verifies at the boundary; the risk still fails and will be reported as an exception.\\

\end{tabular}

\end{table}

\subsubsection{ Step 3 (Output): Validate root posture and issue a posture determination}

Validation occurs once at the root claim. Validation is distinct from verification: verification assesses whether local success rates meet internally configured thresholds τ(n), while validation assesses whether the root claim posture meets externally specified acceptance thresholds (e.g., regulatory expectations or board-approved bands). The output is a posture determination accompanied by a traceable rationale (the chain-of-evidence)\citep{calagna_applying_2021,tabassi_ai_2023,noauthor_regulation_2024}.

\begin{table}
\centering

\begin{tabular}{l}
Running example: root validation produces a unit of trust (with explicit exceptions)

Root posture: s(C\_ex)=s(O\_ex)=0.90.

Root validation: with acceptance bands Green ≥0.90, Amber 0.80–0.90, Red <0.80, the claim validates as Green (at the boundary).

Posture output: (i) root posture value and state, (ii) exception list (R\_ex failed verification), and (iii) evidence pointers (links to the underlying evidence records). \\

\end{tabular}

\end{table}

\subsection{Worked use case: operationalising Australian AI Guardrails for Microsoft 365 Copilot} \label{subsec:taip-co-pilot-example}

To validate the TAIP method beyond theoretical constructs, the study demonstrated an end-to-end execution of the framework within a realistic enterprise environment. The scenario modelled an Australian professional services fictitious firm ("FairOrg")with 2,000 staff that was deploying Microsoft 365 Copilot to a pilot group and had released a custom Copilot Studio agent named 'PolicyHelper'.

The objective of this demonstration was to verify a high-level governance claim derived from the AI Regulation Cluster specifically the Australian Voluntary AI Safety Standard \citep{department_voluntary_2024} by decomposing it into auditable AI System Components and Lifecycle stages \ref{Decomposition AI System Matrix}. This scope is then bound to specific commercial evidence sources (Microsoft Purview, Defender, and PyRIT) \citep{microsoft_learn_2025,microsoft_audit_2025}, to calculate a definitive posture signal.

\begin{figure}[H]
    \centering
    \includegraphics[width=0.75\linewidth]{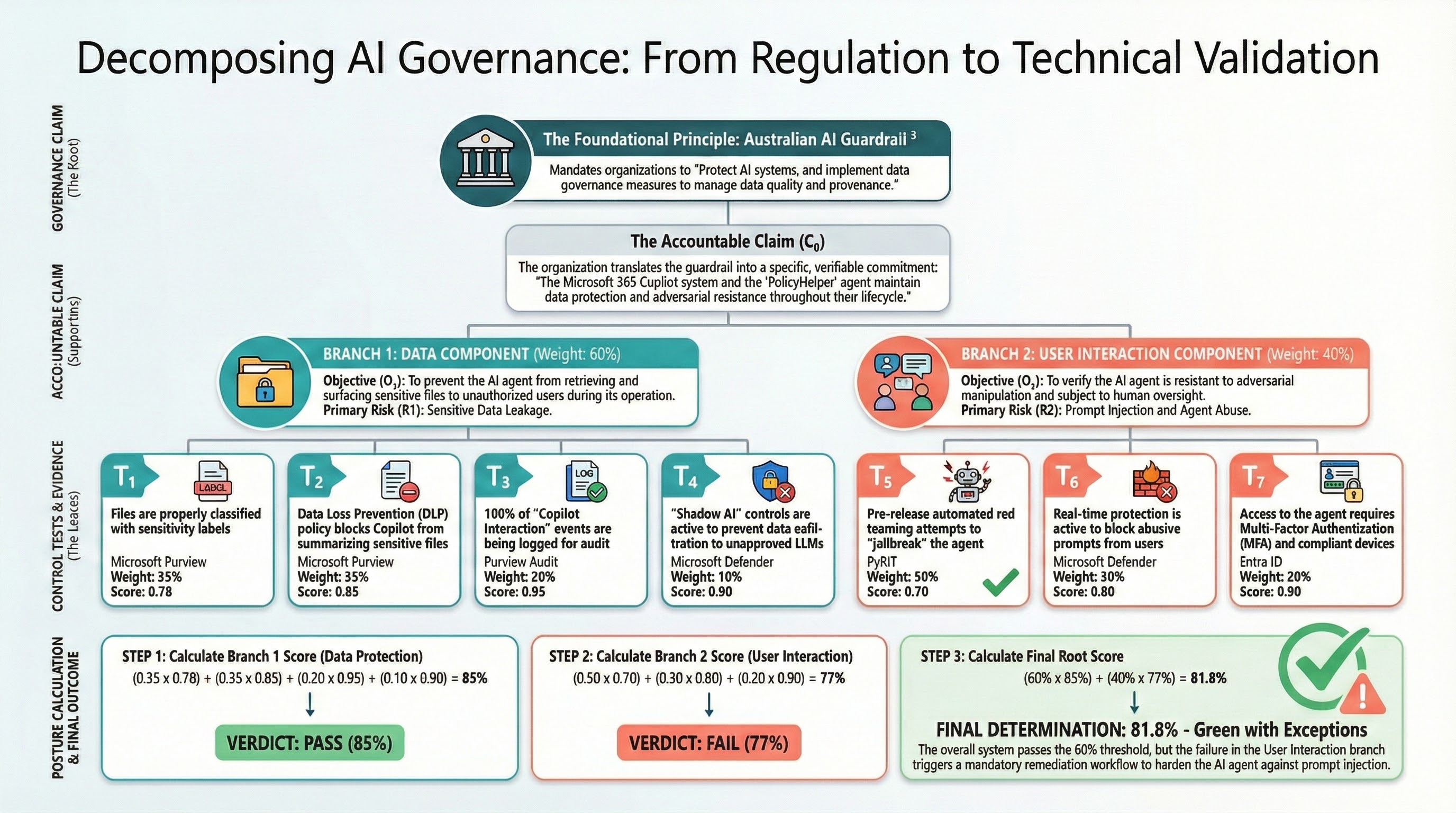}
    \caption{Worked use case posture calculation: mapping Australian AI Guardrail 3 to Microsoft 365 Copilot controls and evidence.}
    \label{fig:Figure10}
\end{figure}
\FloatBarrier

\subsubsection{Input Phase: From regulation to accountable claim}

The workflow initiates in the Accountable Management Cluster, where the organisation's leadership interprets external regulation into a verifiable internal commitment. Guided by Australian AI Guardrail 3, which mandates organisations to "Protect AI systems, and implement data governance measures to manage data quality and provenance", the Board issues a formal accountable claim. This translation step is critical, as effective governance requires converting high-level regulatory principles into specific organisational mandates \citep{mokander_ethics-based_2021-1}.

For this specific deployment, the claim (C0) is formalised as: "The Microsoft 365 Copilot system and the 'PolicyHelper' agent maintain data protection and adversarial resistance throughout their lifecycle."

\subsubsection{Decomposition: Mapping to AI system components and lifecycle}

To transform this high-level claim into executable assurance, the framework decomposes the system boundary using the AI Operations Cluster of the ontology. Rather than assessing "the system" as a monolith, the Scope Construction phase isolates specific AI System Components and their corresponding AI Lifecycle states to pinpoint where risks actually manifest \citep{ieee_7000-2021_2021, markert_gafai_2022}.

For the 'PolicyHelper' agent and Copilot deployment, the profile isolates two distinct operational intersections:

\begin{itemize}
  \item Component A: Data / Lifecycle: Operations. The first decomposition focuses on the Data Component specifically the SharePoint knowledge base accessed by the agent during the Operations (Runtime) lifecycle stage. The AI Assurance Objective (O1) is to prevent the agent from retrieving and surfacing sensitive files to unauthorised users during live interactions, addressing the critical security threat of data leakage in retrieval-augmented generation architectures \citep{chen_security_2024}.
  \item Component B: User Interaction / Lifecycle: Development \& Operations. The second decomposition targets the User Interaction Component. This covers two lifecycle phases: Development (pre-release testing against injection) and Operations (runtime defense against abuse). The AI Assurance Objective (O2) is to verify that the agent is resistant to adversarial manipulation and subject to human oversight, ensuring design-time robustness translates into run-time safety \citep{national_institute_of_standards_and_technology_artificial_2023}.

\end{itemize}
\FloatBarrier

\subsubsection{Risk assignment and evidence binding}

Once decomposed, the framework assigns specific material risks to these intersections and binds them to seven specific control tests (T1--T7).

This binding process treats operational controls as "objects of assurance" that emit normalised evidence records, satisfying the requirement for assurance to be records-based rather than document-based \citep{minkkinen_continuous_2022}.

\paragraph{Branch 1: Data Component (Weight: 0.60)}

The primary risk identified here is R1: Sensitive Data Leakage. To evidence this, the framework binds four specific control tests using Microsoft Purview and Defender:

\begin{itemize}
  \item T1 Label Coverage (Purview): Scans the knowledge base to ensure files are classified (Weight 0.35).
  \item T2 DLP Policy (Purview): Verifies that Data Loss Prevention policies specifically block Copilot from summarising sensitive tags (Weight 0.35).
  \item T3 Interaction Logging (Purview Audit): Verifies that the "Copilot Interaction" audit log is capturing 100\% of events for traceability (Weight 0.20).
  \item T4 GenAI App Control (Defender): Verifies that "Shadow AI" controls are active to prevent data exfiltration to non-corporate LLMs (Weight 0.10).

\end{itemize}
\FloatBarrier

\paragraph{Branch 2: User Interaction Component (Weight: 0.40)}

The primary risk is R2: Prompt Injection and Agent Abuse. This is evidenced by three controls spanning development and operations, aligning with the NIST AI RMF emphasis on testing, evaluation, verification, and validation (TEVV) \citep{national_institute_of_standards_and_technology_artificial_2023,enisa_ai_2023}:

\begin{itemize}
  \item T5 Red Teaming (PyRIT): An automated pre-release test using the Python Risk Identification Tool to attempt jailbreaks (Weight 0.50)
  \item T6 Runtime Protection (Defender): Verifies that the "GenAI Runtime Protection" policy is active to block abusive prompts in real-time (Weight 0.30).
  \item T7 Conditional Access (Entra ID): Verifies that access to the agent requires multifactor authentication and compliant devices (Weight 0.20).

\end{itemize}

\subsubsection{Process Phase: TEV execution and posture calculation}

With the profile configured, the TEV (Test, Evaluate, Verify) engine executes. The engine aggregates the raw evidence from the seven tests using the weighted sum algorithm described in the TAIP method, ensuring risk-informed trust aggregation \citep{fan_decentralized_2021}.

\paragraph{Evaluation of Branch 1 (Data Protection): The operational controls for the data component perform robustly.}
\FloatBarrier
\begin{itemize}
  \item T1 (Label Coverage) scores 0.78 (remediation needed on legacy files).
  \item T2 (DLP) scores 0.85.
  \item T3 (Logging) scores 0.95.
  \item T4 (App Control) scores 0.90.

\end{itemize}
\FloatBarrier
Weighted Calculation for R1:

Verification: Since 0.85$\geq$$\tau$(0.80), the Data Component PASSES.

\paragraph{Evaluation of Branch 2 (User Interaction): The controls for the interaction component reveal specific vulnerabilities, particularly in the automated testing phase.}
\FloatBarrier
\begin{itemize}
  \item T5 (PyRIT Red Teaming) scores 0.70. The automated probe successfully injected prompts in 3 out of 10 attempts.
  \item T6 (Runtime Protection) scores 0.80.
  \item T7 (Access Control) scores 0.90.

\end{itemize}
\FloatBarrier
Weighted Calculation for R2:

Verification: Since 0.77<$\tau$(0.80), the User Interaction Component FAILS verification.

\subsubsection{Output Phase: Posture determination for Accountable Management}

The final phase validates the composite signal against the root claim (C0). Using the composite design pattern inherent to TAIP, low-level technical signals are mathematically rolled up to the governance layer \citep{fan_decentralized_2021}:

Although the specific red-teaming risk (R2) failed, the overall system score of 81.8\% sits above the root acceptance threshold of 80\% defined by the Board.

This results in a "Green with Exceptions" determination.

This structured output allows Accountable Management to validate compliance with Australian Guardrail 3 (Data Governance) based on the strong performance of the Data Component, while the exception on Risk R2 triggers a mandatory remediation workflow to harden the 'PolicyHelper' system instructions against prompt injection. This creates a defensible, evidence-backed position \citep{bloomfield_assessing_2022} where technical failures are visible and actionable, rather than obscured by a binary compliance report.

\subsection{Validation and comparative evaluation strategy} \label{subsec:taip-validation}

TAIP was validated using a dual-lens strategy. First, it was evaluated with the same evidence-gated benchmark used to code the existing framework corpus (see Section~\ref{subsec:rubric-gate} and Appendix~\ref{sec:apdx:b:frameworks}). Second, the Microsoft 365 Copilot demonstration was used to test whether TAIP’s design assumptions hold under real operational conditions (see Section~\ref{subsec:taip-co-pilot-example} and the design principles in Section~\ref{subsec:taip-research-foundation}).

\subsubsection{EQ1: Capability ladder validation (Framework 14)}

To answer Evaluation Question 1 (EQ1), TAIP was coded against the Governance--Operations--Audit (G--O--A) pillars using the benchmark’s non-compensatory constraints (see Section~\ref{subsec:rubric-gate} and Appendix~\ref{sec:apdx:b:frameworks}). The evaluation confirms that TAIP achieves the posture-ready configuration (G4/O4/A4). Comparative results are reported alongside the other frameworks (see Table~\ref{tab:table2}), and the detailed evidence ledger supporting this coding is provided in Appendix~\ref{sec:apdx:b:frameworks}.

\paragraph{Governance Pillar (G):}
TAIP achieves a G4 (posture-ready) rating by defining a reusable and versionable Posture Profile. This profile captures the accountable claim, decomposes it into objectives, and assigns risk-informed thresholds ($\tau$) and weights during scoping (see Step~1 in Section~\ref{subsec:taip-playbook} and the Input--Scope specification in Section~\ref{taip-at-glance-input-scope}). This moves governance from narrative guidance to a structured, repeatable configuration artefact (see the policy abstraction principle in Section~\ref{subsec:taip-research-foundation}). TAIP does not yet reach G5 (ecosystem-ready), because it does not provide a standardised, machine-readable profile schema for cross-organisational sharing and registry-based reuse (see Level~5 requirements in Appendix~\ref{sec:apdx:b:frameworks} and the Level~5 trajectory discussion in Section~\ref{sec:engineering-taip}).

\paragraph{Operations Pillar (O):}
TAIP satisfies O4 (continuous evidence pipeline) by requiring operational controls to emit normalised evidence records that can be ingested continuously. This is specified as a minimal machine-ingestible interface (e.g., timestamp, outcome, metadata) (see the evidence record requirements in Step~2 of Section~\ref{subsec:taip-playbook} and the normalisation principle in Section~\ref{subsec:taip-research-foundation}). The Copilot demonstration shows this binding in practice using heterogeneous sources such as Microsoft Purview and PyRIT (see Section~\ref{subsec:taip-co-pilot-example}). This capability is structurally distinct from the document-centric evidence packs typical of Group A frameworks identified in the comparative morphology (see Section~\ref{subsec:ontology-driven-lit-analysis} and Figure~\ref{fig:Figure4}). The rating is capped at O4 rather than O5 because TAIP defines the interface and binding logic but does not provide a deployable library of connectors/adapters across diverse toolchains (see Level~5 requirements in Appendix~\ref{sec:apdx:b:frameworks} and the Level~5 trajectory in Section~\ref{sec:engineering-taip}).

\paragraph{Audit Pillar (A):}
TAIP achieves A4 (continuous posture monitoring) by defining a repeatable TEVV method that aggregates low-level evidence into higher-level posture signals through explicit roll-up semantics (see the execution workflow in Section~\ref{subsec:taip-playbook} and the posture tree computation logic in Section~\ref{subssec:taip-constructs}). This enables continuous trust signals rather than episodic narrative reports \citep{minkkinen_continuous_2022}. The coding also satisfies the benchmark’s Coupling Constraint C3, which requires A4 to be supported by G4 and O4 (see Appendix~\ref{sec:apdx:b:frameworks}). TAIP is not coded as A5 (agentic audit execution) because, while the method supports automation, it does not yet include an implemented autonomous audit agent capable of self-execution without human orchestration (see the Level~5 trajectory in Section~\ref{sec:engineering-taip} and Appendix~\ref{sec:apdx:b:frameworks}).

\subsubsection{EQ2: Validation of design assumptions against the use case}

Evaluation Question 2 (EQ2) examined whether TAIP’s design assumptions hold under a complex real-world scenario. The Microsoft 365 Copilot demonstration provides affirmative evidence for key assumptions, particularly policy variability, evidence heterogeneity, and traceability (see Section~\ref{subsec:taip-co-pilot-example}).

\paragraph{Policy variability and evidence heterogeneity:}
The Copilot demonstration validated the assumption that the policy content can vary without changing the execution method. The root claim was derived from the Australian AI Guardrails \citep{department_voluntary_2024,industry_safe_2024}, while the TEV execution sequence remained identical to the canonical workflow (see Section~\ref{subsec:taip-playbook}). The same case also confirmed that heterogeneous evidence sources can be bound into a single normalised computation structure, combining structured telemetry (e.g., Purview) and unstructured red-teaming results (e.g., PyRIT) (see the normalised evidence principle in Section~\ref{subsec:taip-research-foundation} and the binding constructs in Section~\ref{subssec:taip-constructs}) (see the policy abstraction pattern in Section~\ref{subsec:taip-research-foundation} and the decoupling rationale in Section~\ref{subsec:taip-at-a-glance}). The same case also \citep{fan_decentralized_2021}.

\paragraph{Traceability and scope dynamics:}
The Copilot demonstration substantiated the assumption that traceability should be generated by the assurance process itself, rather than reconstructed after the fact. In the example, a failure at a leaf node propagated to the root and produced a posture determination of ``Green with Exceptions'' (see Section~\ref{subsec:taip-co-pilot-example}). This shows that the chain-of-evidence is produced deterministically by the posture computation and validation logic (see the recursive TEV flow and validation output in Section~\ref{subssec:taip-constructs}) \citep{bloomfield_assessing_2022}. The case also demonstrated selective scoping, where additional risks and components (e.g., the \textit{PolicyHelper} agent) were incrementally added, consistent with the dynamic boundaries of modern AI deployments \citep{sevilla_compute_2022} (see scoping and revision logic in Step~1 of Section~\ref{subsec:taip-playbook}).

\subsection{Threats to Validity (Executive Summary)}

While the validation is rigorous within the scope of this paper, several limitations should be acknowledged.

\paragraph{1. Benchmark Design Bias}
The capability ladder used to evaluate frameworks was designed by the author. This creates a risk that the structure could unintentionally favour TAIP.

\textbf{Mitigation:}
\begin{itemize}
    \item The ladder uses strict evidence-based scoring rules.
    \item Frameworks must provide tangible artefacts --- not merely advisory language.
    \item Non-compensatory constraints prevent strength in one pillar (e.g., Governance) from masking weakness in another (e.g., Audit).
\end{itemize}

These measures enforce structural coherence rather than subjective maturity scoring.

\paragraph{2. Single Ecosystem Validation}
The worked example is based on a Microsoft 365 Copilot environment. While realistic and enterprise-relevant, it represents one technical ecosystem.

\textbf{Mitigation:}
\begin{itemize}
    \item TAIP is implementation-agnostic.
    \item Policy abstraction is decoupled from execution semantics.
    \item The assurance logic remains invariant across toolchains and regulatory environments.
\end{itemize}

Future research should validate TAIP across open-source LLM chains and fully autonomous multi-agent systems.

\paragraph{3. Limits of Signal Certainty (Epistemic Boundaries)}
TAIP realys on heterogeneous control outputs in the form of success/fail evidence streams. This enables scalability and automation but does not eliminate uncertainty. The posture signal represents confidence within defined thresholds, bounded observation windows, and risk-informed tolerance levels. It does not represent absolute certainty about system behaviour.

\textbf{Mitigation:}
\begin{itemize}
    \item Risk thresholds are explicitly defined by accountable management.
    \item Failures propagate visibly through the posture tree.
    \item Exceptions are preserved even when overall posture validates.
\end{itemize}

Future research may explore probabilistic or uncertainty-aware aggregation models.

\paragraph{4. Evidence Integrity Boundary}
TAIP consumes evidence emitted by operational controls and testing mechanisms. The design, robustness, and adversarial resilience of those controls sit within the AI Operating Environment and are outside TAIP’s scope. TAIP assumes a defence-in-depth architecture in which control testing mechanisms are independently governed and validated.

In this sense, TAIP functions as a governance-layer trust aggregator rather than a control-layer verifier. The posture signal reflects the integrity of the evidence streams provided to it, but does not guarantee that underlying control mechanisms are flawless.

\textbf{Mitigation:}
\begin{itemize}
    \item Explicit evidence binding is required during scoping.
    \item Evidence feeds must be normalised and traceable.
    \item Exception propagation ensures visibility of failure.
\end{itemize}

Future research may incorporate integrity verification mechanisms (e.g., provenance validation or tamper detection) into the aggregation layer.

\paragraph{5. Aggregation Masking Risk}
In any weighted scoring model, there is a risk that local failures could be diluted when aggregated upward.

\textbf{Mitigation:}
\begin{itemize}
    \item Each node is independently verified before aggregation.
    \item Failed risks generate explicit exception records.
    \item Root outcomes may produce ``Green with Exceptions'' rather than binary compliance.
\end{itemize}

Failures remain visible and actionable.

\paragraph{6. Governance vs.\ Automation Balance}
Continuous posture computation could be misinterpreted as replacing governance judgement. TAIP does not remove human oversight.

\textbf{Mitigation:}
\begin{itemize}
    \item Validation occurs explicitly at the root claim level.
    \item Acceptance thresholds are defined by accountable management.
    \item TAIP standardises signal production, not decision authority.
\end{itemize}

Human accountability remains intact.



\section{Conclusions and future research trajectory}
This paper addressed the Agentic AI Assurance Scalability Crisis by identifying a structural gap between evolving governance expectations and existing audit practices. Through ontological analysis and comparative evaluation of current frameworks, it demonstrated that document-centric, checklist-based assurance models are not architecturally suited to large-scale, agentic AI environments.
The proposed Trustworthy AI Posture (TAIP) framework introduces a method-level innovation: the decoupling of policy requirements (“what”) from assurance execution logic (“how”). By operationalising a reusable TEVV-based assurance pattern and embedding it within a compositional posture architecture, TAIP enables scalable trust signal generation across heterogeneous technical and governance environments. The dual-lens validation   benchmark evaluation and applied case demonstration   confirms both structural soundness and practical feasibility.
Future research will extend TAIP beyond organisational deployment toward ecosystem-level trust infrastructures. This includes developing executable reference implementations, code samples, and automation artefacts to support machine-speed assurance. A key direction is the development of industry-aligned Posture Cards   standardised posture tree configurations for sectors such as healthcare and telecommunications. When combined with the TAIP framework, such shared structures may provide the foundation for collaborative, interoperable trust mechanisms and contribute to a broader “horizon of trustworthiness” across AI ecosystems.

\section{Artefact availability}

Supplementary artefacts required to reproduce the method are included in the appendices: the Trustworthy AI Assurance Ontology cluster summaries (Appendix A) and the evidence-gated capability ladder rubric and benchmark results (Appendix B). The worked use case in Section 3.6 provides an end-to-end example of claim calibration and evidence binding.

\appendix

\section{The Trustworthy AI Assurance Ontology}
\label{sec:apdx:a:TAI-ontology}

This appendix details the Trustworthy AI Assurance Ontology, the semantic instrument developed to structure the research and the TAIP framework. The ontology harmonizes the vocabulary, information flows, and dependencies required to generate a valid trustworthiness signal. It is composed of six interlinked clusters that model the flow from regulatory obligation to operational evidence.

\subsection{AI Regulations Cluster}

Definition: This cluster forms the normative foundation of the ontology. It captures the external rules and soft-law instruments that shape the definition of responsible AI.

Key nodes: Regulations; voluntary principles; standards; best-practice guidance; sectoral and cross-sector classification schemes.

Key relations: Regulations constrain and guide responsible AI practices; regulations set boundaries for accountable management and inform the selection and interpretation of assurance frameworks.

Evidence: \citep{adedokun_global_2024,industry_safe_2024,boza_implementing_2021,centre_guidelines_nodate,diaz-rodriguez_connecting_2023,ebers_standardizing_2022,floridi_capai_2022,google_secure_2023,hagendorff_blind_2022,high-level_expert_group_on_artificial_intelligence_ai_hleg_ethics_2019,huang_overview_2022,noauthor_implementing_nodate,isoiec_iso_2022,iso_iso_2023,isoiec_iso_2022,noauthor_oecd_2022,noauthor_regulation_2024,lu_responsible_2024,lucaj_ai_2023,luetge_german_nodate,mittelstadt_principles_2019,mokander_us_2022,mokander_ethics-based_2021-1,morley_ethics_2021,prem_ethical_2023,singh_fair_2022,stettinger_trustworthiness_2024,worsdorfer_bidens_2024}.

\subsection{Accountable Management Cluster}

Definition: This cluster anchors trust in organisational governance. It models how leadership interprets regulatory duties and converts them into formal commitments (Claims) that the system must satisfy.

Key nodes: Accountable management; board and executive oversight; governance claims; policies and mandates; risk appetite and tolerances.

Key relations: Accountable management empowers and obligates governance bodies; governance bodies issue claims and approve policies; claims define the objectives and acceptance criteria that downstream GRC and audit must satisfy.

Evidence: \citep{agbese_governance_2023,akula_audit_2021,noauthor_implementing_nodate,baker-brunnbauer_trustworthy_2021,baquero_derisking_nodate,beranger_societal_2021,brundage_toward_2020,calagna_applying_2021,diaz-rodriguez_connecting_2023,european_union_agency_for_cybersecurity_artificial_2023,floridi_capai_2022,javadi_monitoring_2020,us_government_accountability_office_gao-21-519sp_2021,hickman_trustworthy_2021,high-level_expert_group_on_artificial_intelligence_ai_hleg_ethics_2019,lee_responsible_2025,lu_towards_2022,luciano_capai-procedure_2022,mantymaki_putting_2023,markovic_accountability_2021,mcgrath_enterprise_nodate,mylrea_ai_2023,mokander_us_2022,mokander_ethics-based_2021-1,verborgh_semantic_2021,novelli_accountability_2024,pwc_responsible_2020-1,raja_ai_2023,national_institute_of_standards_and_technology_artificial_2023,department_voluntary_2024}.

\subsection{AI Operations Cluster}

Definition: The operational core where the AI system is built, deployed, and monitored. This cluster is the source of all technical evidence.

Key nodes: AI system components (data, model, infrastructure, API); lifecycle stages (design, development, deployment, operations, retirement); operational controls; telemetry and logs; evidence records.

Key relations: Operations implement controls and generate evidence; telemetry/logs are normalised into evidence records; evidence is consumed by audit and aggregated into posture signals.

Evidence: \citep{ieee_7000-2021_2021,isoiec_iso_2022,noauthor_mitre_nodate,noauthor_oecd_2022,google_secure_2023,agarwal_seven-layer_2024,agbese_implementing_2023,alfrink_contestable_2022,avinash_seven-layer_2022,banu_preventing_2020,beranger_societal_2021,boza_implementing_2021,brundage_toward_2020,clement_xair_2023,comission_assessment_2020,enisa_artificial_2023,etzioni_designing_2016,sison_chatgpt_2024,google_secure_2023,hallensleben_principles_2020,hu_artificial_2023,hupont_landscape_2022,isaca_auditing_2018,khlaaf_toward_2023,li_trustworthy_2023,lu_towards_2022,mantymaki_putting_2023,markovic_accountability_2021,mckinsey_derisking_2020,mylrea_ai_2023,morley_ethics_2021,verborgh_semantic_2021,solanki_operationalising_2023,national_institute_of_standards_and_technology_artificial_2023,vakkuri_how_2022,wei_trustworthy_2025,xu_generative_2024}.

\subsection{AI Audit Cluster}

Definition: The verification layer that consumes operational evidence to validate governance claims. It defines the logic of the "Test, Evaluate, Verify, Validate" (TEVV) cycle.

Key nodes: Audit process; control tests; risk evaluation; verification against policy/thresholds; validation against external acceptance criteria.

Key relations: Audit consumes operational evidence; tests and evaluation verify controls and policy alignment; validation determines whether the root claim is acceptable given risk appetite and external criteria.

Evidence: \citep{ieee_7000-2021_2021,iso_iso_2022,isoiecieee_iso_2022,isoiecieee_iso_2022,akula_audit_2021,boer_algorithm_2023,alfrink_contestable_2022,noauthor_implementing_nodate,baquero_derisking_nodate,beranger_societal_2021,berghout_advanced_2023,calagna_applying_2021,costanza-chock_who_2022,european_union_agency_for_cybersecurity_artificial_2023,fukas_developing_2021,us_government_accountability_office_gao-21-519sp_2021,isaca_auditing_2018,isoiec_iso_2023,konigstorfer_ai_2022,lam_framework_2024,lee_responsible_2025,li_making_2024,luciano_capai-procedure_2022,markert_gafai_2022,mokander_auditing_2024,mokander_ethics-based_2021-1,ojewale_towards_2024,pal_implementing_2021,national_institute_of_standards_and_technology_artificial_2023,toader_auditability_2019,uk_kpmg_2018,wong_seeing_2023,xia_towards_2023}.

\subsection{AI Governance, Risk and Compliance (GRC) Cluster}

Definition: The translation layer that converts high-level claims into enforceable policies, risks, and thresholds.

Key nodes: Risk management; risk register and treatment plans; governance policies; AI Assurance Objectives; risk appetite thresholds; compliance reporting.

Key relations: Claims are decomposed into objectives and material risks; risk appetite defines tolerances and expected success thresholds; objectives drive the selection of controls/tests and govern how evidence is interpreted.

Evidence: \citep{kaur_trustworthy_2023,lu_software_2022,luciano_capai-procedure_2022,wong_seeing_2023,zinda_ethics_2022}.

\subsection{Trustworthy AI Frameworks Cluster}

Definition: This cluster structures the methodology. It represents the standardized approaches organisations adopt to execute the assurance process.

Key nodes: Trustworthy AI frameworks; audit/assurance and risk-management methodologies; supporting artefacts (standards, templates, design patterns, best practices).

Key relations: Frameworks operationalise governance, operations, and audit mechanisms; regulations and governance claims shape framework requirements; frameworks specify artefacts and processes used to produce and interpret evidence.

Evidence: \citep{akula_audit_2021,jobin_global_2019,baker-brunnbauer_trustworthy_2021,boza_implementing_2021,commission_assessment_2020,de_almeida_artificial_2021,raji_closing_2020,diaz-rodriguez_connecting_2023,dignum_responsible_2019,european_union_agency_for_cybersecurity_artificial_2023,falco_governing_2021,high-level_expert_group_on_artificial_intelligence_ai_hleg_ethics_2019,iia_iias_2018,morley_what_2021,lu_responsible_2024,luciano_capai-procedure_2022,mantymaki_putting_2023,markert_gafai_2022,mittelstadt_principles_2019,mylrea_ai_2023,novelli_accountability_2024,pelosi_hybrid-dlt_2023,raja_ai_2023,schlicker_how_2022,national_institute_of_standards_and_technology_artificial_2023,vakkuri_how_2022,wickramasinghe_trustworthy_2020,zicari_z-inspection_2021}.



\section{Selected Frameworks, and Gated Rubric}

\subsection{Frameworks}
\label{sec:apdx:b:frameworks}

Frameworks included in the benchmark are listed in Table \ref{13-framework-table}.

\subsection{Gated Rubric}

To rigorously evaluate whether existing frameworks and the proposed Trustworthy AI Posture (TAIP) could support the demands of the Large-Scale Era \citep{sevilla_compute_2022, mokander_auditing_2024, jaffri_hype_2023} , this study operationalised the Trustworthy AI Assurance Ontology as an evidence-based assessment instrument: the Capability Ladder. The Capability Ladder design, deterministic coding rules applied to the corpus, and comparative results for the thirteen state-of-the-art frameworks and the proposed TAIP solution are presented in the subsections that follow.

\subsection{The Capability Ladder Rubric}

Unlike subjective maturity models that rely on self-assessment, this ladder uses a binary, evidence-gated rubric. It evaluates frameworks based on their structural capacity to support continuous assurance. The ladder is organised into three interdependent pillars derived from the ontology: AI Governance (G), AI Operations (O), and AI Audit (A).

\subsubsection{Maturity Levels (0--5)}

Each pillar is assessed on an ordinal scale ranging from Ad-hoc (Level 0) to Agentic (Level 5). In the context of this paper, Level 4 (Posture-Ready) is established as the required benchmark for continuous, automated assurance. Level 5 represents a future horizon of ecosystem interoperability.

\begin{itemize}
    \item Level 0 (Ad-hoc): Processes are undefined; assurance is implicit.
    \item Level 1 (Document-Centric): Checklists and principles exist but rely on static documentation (e.g., PDF forms).
    \item Level 2 (Structured/Control-Aware): Risks and controls are structured, but evidence collection remains manual.
    \item Level 3 (Machine-Ingestible): Frameworks define schemas or APIs for evidence, enabling partial automation.
    \item Level 4 (Posture-Ready): The target benchmark. Requires reusable governance profiles, continuous evidence pipelines, and automated verification logic.
    \item Level 5 (Agentic/Ecosystem): Fully autonomous audit agents and cross-organisational profile sharing.
\end{itemize}

\subsubsection{Evidence Admissibility Rules}

To ensure objectivity, the coding process is governed by strict admissibility rules (R-E1 to R-E6). The most critical of these is the Provision vs. Requirement Rule (R-E2). A framework is not credited with a capability simply because it states an organisation should perform a task (e.g., "maintain a risk register"). To score, the framework must provide the artifact, schema, or template to execute that task. The Tool Mention Rule (R-E3) dictates that vague references to "automation" do not qualify; the framework must specify the interface or logic used for that automation.

\subsection{Qualifier Questions and Coupling Constraints}

The ladder assigns maturity levels based on a set of 22 binary qualifier questions across the three pillars. To prevent structural inconsistencies such as claiming continuous auditing without continuous data Coupling Constraints are applied to the final score.

\subsubsection{Pillar Qualifiers (Summary)}

\begin{itemize}
      \item Governance (G) -- 7 Qualifiers: These questions assess whether the framework provides reusable scope artefacts, explicit assurance claims, structured risk taxonomies, and defined performance thresholds. The highest levels test for the existence of versionable, machine-readable Posture Profiles (GQ6).
      \item Operations (O) -- 7 Qualifiers: These assess the nature of evidence generation. Lower levels check for document-based evidence. Higher levels (O3--O4) test for the specification of machine-ingestible interfaces (APIs, schemas) and continuous telemetry pipelines.
      \item Audit (A) -- 8 Qualifiers: These assess the execution logic. Questions determine if the audit is episodic or continuous, narrative or signal-based. Level A4 requires the definition of roll-up semantics that mathematically aggregate evidence into a posture signal (AQ7).
\end{itemize}
  
\subsubsection{Coupling Constraints (Non-Compensatory Logic)}

High capability in one pillar cannot compensate for deficits in another. The coding logic enforces dependencies through constraints:

\begin{itemize}
    \item Constraint C1 \& C2: High-level audit execution is impossible without structured governance and operational evidence. If G or O are low, the Audit score is capped at A2.
    \item Constraint C3 (The Posture Constraint): A framework cannot be coded as Continuous Posture Monitoring (A4) unless it simultaneously possesses Posture-Ready Governance (G4) and Continuous Evidence Pipelines (O4). This reflects the reality that one cannot automate the verification of a policy that does not exist or evidence that is not there.
\end{itemize}

\subsection{Benchmark Analysis and Validation Results}

The rubric was applied to the thirteen frameworks identified in the Systematic Literature Review (Section 2) and to the TAIP framework (Section 3) as a validation exercise.

\subsubsection{The Structural Gap in Existing Frameworks}

The analysis revealed a distinct morphology in the current landscape. Frameworks tended to cluster into Group A (Process-Heavy), which define rigorous manual workflows (e.g., CapAI, SMACTR), or Group B (Principle-Heavy), which focus on ethical definitions but lack execution logic (e.g., ALTAI, NIST AI RMF). A third cluster, Group C (Technical), offers continuous evidence mechanisms (e.g., Hybrid-DLT) but lacks the governance layer to interpret that data.

None of the thirteen existing frameworks achieve the Level 4 benchmark. They either lack the machine-readable governance profiles or the continuous evidence aggregation logic required for the large-scale era.

\subsubsection{Validation of TAIP (Framework \#14)}

TAIP was evaluated as Framework \#14 using the same evidence-gated strictures. The analysis confirms that TAIP closes the structural gap.

\begin{itemize}
    \item G4 (Governance): TAIP provides the Posture Profile, a reusable, versionable schema for defining claims, risks, and thresholds (satisfying GQ6).
    \item O4 (Operations): TAIP defines a normalised, machine-ingestible Evidence Record schema and continuous pipelines (satisfying OQ4 and OQ6).
    \item A4 (Audit): TAIP defines the TEV (Test-Evaluate-Verify) Engine, which provides the computational roll-up logic to convert evidence into a posture signal (satisfying AQ7).
\end{itemize}

Table 3 summarises the final classification of the full corpus, demonstrating that TAIP is the sole framework to achieve the G4/O4/A4 configuration required for scalable, continuous assurance.

\begin{table}[H] 
  \centering   
  \begin{tabularx}{\linewidth}{@{} >{\raggedright\arraybackslash}X c c c c c @{}}
    \toprule
    \textbf{Framework} & 
    \textbf{G} (Gov) & 
    \textbf{O} (Ops) & 
    \textbf{A} (Audit) & 
    \textbf{Ready?} & 
    \textbf{Group} \\
    \midrule
    Ethics-Based Auditing (Mökander et al.) & 1 & 1 & 0 & No & B \\
    TAII Framework (Baker-Brunnbauer) & 0 & 0 & 0 & No & B \\
    SMACTR (Raji et al.) & 1 & 2 & 2 & No & B \\
    GAFAI (Markert et al.) & 3 & 2 & 2 & No & A \\
    CapAI (Floridi et al.) & 3 & 2 & 2 & No & A \\
    Z-Inspection\textregistered\ (Zicari et al.) & 1 & 1 & 1 & No & A \\
    ALTAI (EU HLEG) & 1 & 1 & 0 & No & B \\
    NIST AI RMF 1.0 (NIST) & 2 & 1 & 0 & No & B \\
    ITAF 4th Ed. (ISACA) & 0 & 0 & 2 & No & A \\
    Algorithm Assurance (de Boer et al.) & 1 & 1 & 1 & No & A \\
    SAO Framework (Naja et al.) & 0 & 3 & 0 & No & C \\
    Assurance Audits (Lam et al.) & 2 & 1 & 2 & No & A \\
    Hybrid-DLT Framework (Pelosi et al.) & 0 & 4 & 0 & No & C \\
    \midrule
    \textbf{TAIP (Proposed Framework)} & \textbf{4} & \textbf{4} & \textbf{4} & \textbf{YES} & \textbf{C} \\
    \bottomrule
  \end{tabularx}
  \caption{Comparative classification of Trustworthy AI frameworks (capability ladder results).}
  \label{tab:table2}
\end{table}

Note: Level 4 (G4/O4/A4) represents the minimum configuration for continuous, automated posture. Level 5 (Ecosystem/Agentic) was not achieved by any framework in the current corpus.

\printbibliography

\end{document}